\documentclass[a4paper,11pt]{article}
\usepackage{jheppub} 

\usepackage{microtype} 

\usepackage{amsmath,amstext, amssymb, amsthm, amsfonts, slashed}
\usepackage{physics}
\usepackage{mathtools}
\usepackage[version=4]{mhchem}
\usepackage{graphicx}
\usepackage{booktabs}
\usepackage{dcolumn}
\usepackage{bm}
\usepackage{bbm}
\usepackage{dsfont}
\usepackage[usenames,dvipsnames]{xcolor}
\usepackage[normalem]{ulem}
\usepackage{footmisc}

\usepackage{url}
\usepackage[colorlinks=true,urlcolor=blue,linkcolor=blue,citecolor=blue,anchorcolor=blue,filecolor=blue,menucolor=blue,linktocpage=true,pdfproducer=medialab,pdfa=true]{hyperref}
\usepackage[nameinlink]{cleveref}
\usepackage{braket}
\usepackage{simplewick}
\usepackage{float}
\usepackage{comment}

\newcommand{\supsetsim}{\mathrel{\ooalign{\raise.4ex\hbox{$\supset$}\cr$\raise-.9ex\hbox{$\sim$}$}}}

\hypersetup{
	colorlinks,
	linkcolor={blue!75!black},
	citecolor={blue!75!black},
	urlcolor={blue!75!black}
}

\makeatletter\gdef\@fpheader{}\makeatother 
\newcommand{\orcid}[1]{\href{https://orcid.org/#1}{	\raisebox{0.5\height}{\includegraphics[height=1.25ex,width=1.25ex]{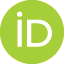}}}}

\newcommand{\kSSBone}{k_{\textrm{SSB},1}}

\newcommand{\LUV}{\Lambda_{\textrm{UV}}}

\newcommand{\commentmute}[1]{} 

\graphicspath{{Figures/}}

\begin{document}

\title{The~quantum~criticality~of~the~Standard~Model and~the~hierarchy~problem}

\author[a]{Juan~Pablo~Garc\'es,\orcid{0000-0002-6933-8750}\,}
\author[a]{Florian~Goertz,\orcid{0000-0001-8880-0157}\,}
\author[a]{Manfred~Lindner,\orcid{0000-0002-3704-6016}\,}
\author[b,a,c]{and \'Alvaro~Pastor-Guti\'errez\orcid{0000-0001-5152-3678}\,}
\affiliation[a]{Max-Planck-Institut f\"ur Kernphysik P.O. Box 103980, D 69029, Heidelberg, Germany}
\affiliation[b]{RIKEN Center for Interdisciplinary Theoretical and Mathematical Sciences (iTHEMS), RIKEN, Wako 351-0198, Japan}
\affiliation[c]{Institut für Theoretische Physik, Universität Heidelberg, Philosophenweg 16, 69120 Heidelberg, Germany}

\preprint{RIKEN-iTHEMS-Report-25}

\abstract{
The naturalness principle has long guided efforts to understand physics beyond the Standard Model, with the hierarchy problem as the central issue. We revisit the role of quantum corrections in the fine-tuning of the low-energy effective description and its phase structure. We implement, for the first time in this context, the full Standard Model within the Wilsonian functional renormalization group. Crucially, this method captures conveniently both logarithmic and quadratic scalings, which must both be considered in the tuning, and allows us to provide a new generic and quantitative study of fine-tuning and its interpretation in terms of critical phenomena. We emphasize on the connection between the hierarchy problem and the near-criticality of the Standard Model and extract scheme-independent information on the infrared Higgs phases and the associated quantum phase transition as well as discuss a related enhanced fine-tuning usually not considered in tuning estimates. Finally, we illustrate the framework’s versatility by exploring new physics coupled to the Higgs sector that can soften high-scale sensitivity, recovering also the large-anomalous-dimension solution to the hierarchy problem.
}

\maketitle

\section{Introduction}
So far, the Standard Model (SM) of particle physics has been remarkably successful in describing nature at the most fundamental level accessible to experiment. Despite this success, established evidence shows that the SM cannot be the ultimate theory but rather a highly accurate low-energy limit. It fails to explain the origin of dark matter, neutrino masses, and the baryon asymmetry of the universe, indicating the need for physics beyond the SM (BSM).

Many extensions of the SM have been proposed to address these shortcomings. Since no breakdown of the SM is observed below the Planck scale, such extensions are also motivated by internal \textit{puzzles} of the SM itself, such as the fermion mass hierarchies or the absence of CP violation in strong interactions. A particularly intriguing puzzle is the origin of the electroweak (EW) scale, set by the Higgs mass or vacuum expectation value, and its relation to possible new high-energy scales. The strong dependence of infrared (IR) physics on ultraviolet (UV) embeddings, rooted in the single dimensionful parameter of the SM (the Higgs mass), constitutes the \textit{hierarchy problem}. Such sensitivity is encoded in quantum corrections to correlation functions -- here the Higgs two-point function. Canonical power counting implies that dimensionful quantities receive two kinds of corrections: logarithmic and power-law. The tension arises in the presence of the latter and, specifically, from the interplay of both types. Obtaining the correct IR physics then requires fine adjustment of UV parameters, in conflict with the principle of naturalness~\cite{tHooft:1979rat,Giudice:2008bi,Dirac:1938mt,Craig:2022eqo}. While not a logical inconsistency that must be solved (unlike, e.g. dark matter), naturalness provides a guiding criterion for viable theoretical scenarios.

In particular, the absence of supersymmetry at current collider energies has renewed interest in alternative approaches to address the hierarchy problem. Experimentally, the lack of new physics up to multi-TeV scales points to a \textit{little hierarchy} problem, motivating mechanisms that provide a natural extension and raise the expected new-physics scale. Examples include Little Higgs models~\cite{ArkaniHamed:2001nc,ArkaniHamed:2002qx,ArkaniHamed:2002qy,Low:2002ws,Ahmed:2023qsm}, where the Higgs emerges as a pseudo–Goldstone boson, lifting the scale by roughly an order of magnitude, as well as TeV-scale hidden sectors~\cite{Chacko:2005pe,Batell:2022pzc,Bally:2022naz,Chung:2023gcm}. In this context, considering the little hierarchy as a genuine result of a single more fundamental model, rather than an independent issue, might lead to comprehensive solutions, see e.g. the recent~\cite{Blasi:2019jqc,Blasi:2020ktl,Blasi:2022hgi,Angelescu:2023usv,Chung:2023gcm,Ahmed:2023qsm,Cacciapaglia:2020kgq,Cacciapaglia2022a}.

For a comprehensive understanding of the problem, which may potentially motivate the study of unexplored scenarios, the power-law corrections to the Higgs mass, being central to the UV–IR sensitivity, need to be analyzed in detail. They scale with the square of a given UV cutoff, which marks the limit of validity of the effective theory. This issue was first highlighted in studies of Grand Unification Theories (GUTs), where the Higgs mass had to be finely tuned relative to the GUT scale, giving rise to the \textit{gauge hierarchy problem}\cite{Gildener1976,Weinberg1979a}. In standard perturbative approaches, these corrections only emerge at thresholds because of the mass-independent nature of the regularization. Consequently, to access these power-law contributions with these methods, one must postulate a specific UV completion. In contrast, Wilsonian renormalization schemes~\cite{Wilson:1974mb}, which use a cutoff to progressively integrate quantum fluctuations, naturally account for power-law corrections. In this context, the UV cutoff can be interpreted as the result of integrating out a more fundamental embedding. This bottom-up approach does not require specifying a UV completion to manifest this polynomial sensitivity to high scales.

In this work, we investigate the role of corrections to the Higgs mass in the context of quantum criticality and naturalness in the SM using the functional Renormalization Group (fRG)~\cite{Wetterich:1992yh}. The fRG is a quantitative and robust method that has been successfully applied across a wide range of fields, from condensed matter, confining gauge theories, and dynamical chiral symmetry breaking to quantum gravity; see~\cite{Dupuis:2020fhh} for a review. In high-energy particle physics, including the SM and its extensions, the default approach relies on perturbative methods. However, the fRG has been used to study individual sectors such as the Higgs~\cite{Gies:2013fua,Gies:2014xha,Eichhorn:2015kea,Reichert:2017puo,Sondenheimer:2017jin,Gies:2017ajd,Held:2018cxd,Eichhorn:2020upj}, Yukawa~\cite{Borchardt:2016xju,Gies:2017zwf,Ellwanger:1992us}, and gauge-Yukawa~\cite{Gies:2019nij,Gies:2013pma,Gies:2018vwk,Gies:2016kkk,Gies:2015lia,Gies:2023jzd} sectors. The present work builds on the full implementation of the SM developed in~\cite{Pastor-Gutierrez:2022nki,Goertz:2023pvn}, which consistently includes all SM sectors from Planckian to sub-QCD scales.

Despite its structural advantage in exposing power-law sensitivity, the Wilsonian approach has only been applied in a few studies of naturalness~\cite{Krajewski:2014vea,Krajewski:2015eea,Gies:2023jzd}.
The first study in this direction was performed in~\cite{Krajewski:2014vea,Krajewski:2015eea} for a Higgs-Yukawa toy model, exploring the fine-tuning required as a function of the new-physics scale and its implications for vacuum stability. More recently, in~\cite{Gies:2023jzd}, the interplay between different sources of chiral symmetry breaking in the SM - through Higgs and emergent condensates - was studied in detail, with a special focus on its impact on the quantum phase transition. There, it was shown that within the SM class of theories, there exists a natural limit where fine-tuning is minimized as the QCD and EWSB scales become similar. These studies highlight the strength of Wilsonian methods in assessing more natural configurations in theory space.

Here, we add to previous studies by analyzing the near-critical nature of the SM, understood as its proximity to an EW-symmetric solution, and its relation to the hierarchy problem in the context of new physics. To this end, we exploit the analogy between the IR cutoff and a physical temperature to investigate the complete phase structure of SM-like theories along the RG flow. In addition, we study the fine-tuning problem in the SM effective theory from a bottom-up perspective and explore how BSM deformations in theory space affect it. This provides a simplified but systematic tool for the search for new physics based on the principle of naturalness, which is particularly relevant in light of the absence of new physics at scales expected from naturalness arguments. The following analysis is, to the best of our knowledge, the first fRG-based naturalness study that incorporates the full Standard Model and provides quantitative results.

This work is organized as follows. First, in \Cref{sec:functionalSM}, we introduce the SM results obtained from functional methods and place them in the context of the standard perturbative picture. This comparison allows us to identify what information is already encoded in the Wilsonian scheme, such as the evolution across different Higgs phases. We then proceed in \Cref{sec:SMPT} to analyze the phase structure of the general SM class of theories and its evolution along the RG flow, particularly focusing on the scalar sector. Having set the stage, in \Cref{sec:quantumPT}, we study the quantum phase transition present in the class of theories and its close relation to the hierarchy problem, highlighting the impact of the near-criticality of the SM even when new physics enters at the TeV scale, which is particularly important in the context of the little hierarchy problem. In \Cref{sec:naturalness}, we quantify the fine-tuning associated with deforming the SM in various directions in parameter space, and use this to search for new physics scenarios that minimize fine-tuning in \Cref{sec:newPhysicsdeformations}. Finally, we conclude in \Cref{sec:conclusions}.

\section{The Standard Model from a functional point of view}\label{sec:functionalSM}
In this Section we commence by briefly reviewing the flow equations for the SM, paying particular attention to the Higgs curvature mass, which is the parameter encoding the information about the phases of the theory and the leading source of sensitivity to UV scales. Our first goal is to introduce the notion of RG flow and the appearance of the SM within this threshold-sensitive and mass-dependent RG scheme. Second, we explain how information about the Higgs-potential phase evolution can already be extracted qualitatively in the present vacuum computation.

The fRG is a Wilsonian approach that systematically accounts for quantum fluctuations by integrating out momentum shells. This is achieved using a regulator function that suppresses infrared modes in the path integral measure below a cutoff scale $k$. This line offers several advantages relevant to our study: first, it provides a finite, mass-dependent renormalization scheme that naturally resolves near-threshold physics and second, it accounts for power-law contributions along the RG flow. As we will make apparent, these contributions are crucial in phase transitions and help differentiate between distinct phases of a theory. This constitutes a qualitative improvement over conventional mass-independent regularization schemes~\cite{tHooft:1972tcz,Bollini:1972ui,Fujikawa:2011zf}, particularly in analyzing the Higgs phase diagram. Additionally, power-law contributions play an essential role in understanding the relationship between different physical scales, which is central to the gauge hierarchy problem. 

In the present analysis, we consider a polynomial-type Higgs potential expanded up to the first marginal coupling,
\begin{align}\label{eq:ueff}
	u_k(\bar\rho ) =V_{\text{eff},\,k}(\rho)\,/ \,k^4= \bar \mu^2_k \, \bar \rho +\bar \lambda_k \,\bar \rho^2 \,,
\end{align}
where
\begin{align}\label{eq:barrho}
	\bar\rho &=   \frac{ \tr \Phi^\dagger \Phi}{k^2}\,, 
	&\text{with}&
	&\Phi= \frac{1}{\sqrt{2}}\begin{pmatrix} Z_{\mathcal{G}_1,k} \, \mathcal{G}_1 +i  Z_{\mathcal{G}_2,k} \, \mathcal{G}_2 \\   Z_{H,k}\,  H  +i Z_{\mathcal{G}_3,k}\, \mathcal{G}_3  \end{pmatrix} \,,
\end{align}
and $H,\, {\cal G}_{1,2,3}$ being the Higgs and Goldstone real scalar fields, respectively, and $Z_{\Phi_i,k}$ accounts for the wave function renormalization of the different modes in the scalar doublet. The scale dependence of dressings and couplings will, from now on, be understood and we will only invoke it in the subscript when referring to a particular scale. 

The effective potential in \eqref{eq:ueff} is parametrized with the dimensionless and renormalized Higgs curvature mass $\bar \mu^2 =\mu^2 k^{-2}$ and quartic self-coupling $\bar \lambda$. It is important to note that the effective potential includes an infinite number of operators and the form specified above neglects the canonically irrelevant ones. These higher dimensional operators can be understood as a parameterization of the field dependence of the relevant and marginal couplings and hence carry information of the large field curvature of the Higgs potential. This is relevant to describe the global properties of the potential such as those concerning metastability. More accurate potentials have be considered in the literature, see eg.~\cite{Reichert:2017puo,Gies:2017ajd,Gies:2017zwf,Borchardt:2016xju} for non-trivial expansions. For the study of the phase diagram of the SM as a low-energy effective theory (EFT), higher-dimensional operators are expected to bring no qualitatively new information given the weak-coupling nature of the theory. However, in certain models the effect of integrating out new physics could source them non-negligibly, complicating the analysis. For the sake of simplicity, we have neglected such effects here. 

The scale-dependent effective potential is expanded around its minimum value $\bar \rho_0$, at which the minimum condition,
\begin{align}\label{eq:minimumderivation}
	\left.	\partial_{\bar \rho}\, u (\bar\rho)\right|_{\bar\rho_0}=0=  \bar \mu^2  +  2\, \bar \lambda  \bar  \rho_0\,,
\end{align}
is satisfied. As will become apparent when we integrate the flow of the effective potential, we can extract useful information concerning the phase structure of the theory from the shape of $u(\bar \rho)$ at finite values of the cutoff $k$. With this we can trace the evolution of the potential between different phases and extract physical information. For cutoff values where $\bar\mu^2>0$ and $\bar \lambda>0$ the effective average potential displays a trivially stable shape and is expanded around $\bar\rho_0=0$. In contrast, for cut-off scales where $\bar\mu^2<0$, the potential displays a nontrivially stable shape with a minimum at
\begin{align}\label{eq:rho0v}
\bar\rho_0= \bar v^2/2 =-\bar\mu^2/ (2 \, \bar \lambda )\geq 0
\end{align}
where $\bar v=  v\, k^{-1}$ is the dimensionless and renormalized Higgs vacuum expectation value. In this regime, the EW symmetry is broken, and the potential is expanded around the non-trivial minimum. Here, it is more convenient to parameterize the potential in terms of $\bar\rho_0$ and $\bar \lambda$, without loss of generality. Furthermore, we define the Euclidean mass of the Higgs boson, which is given by the small-field curvature of the potential around the minimum 
\begin{align}\label{eq:EuclideanmH}
m^2_\textrm{H}= k^2 \left[ \partial_{\bar \rho} u(\bar \rho)+2 \bar \rho \partial^2_{\bar \rho} u(\bar \rho) \right]_{\bar \rho =\bar \rho_0}=\left\{ 
\begin{array}{rcl} 
	 \mu^2 & \hspace{2cm} & \bar \rho_0 =0\\[1ex] 
	 2 \bar \lambda v^2 & & \bar \rho_0 >0
\end{array} \right.\,.
\end{align}
This quantity is, up to momentum-dependent corrections, very close to the pole mass in the $k\to0$ limit.

The flow of the scalar sector parameters can be extracted by performing field derivatives of the flow of the effective potential. Although its derivation is simple, here we omit the details and refer to~\cite{Goertz:2023pvn,Pastor-Gutierrez:2022nki,PastorGutierrez:2025zvx} for a detailed description. The flow of the Higgs curvature mass reads in the symmetric phase,
\begin{align}\label{eq:flowmu}
\left.	\partial_t \,\bar \mu^2\right|_{\bar \rho_0=0}&=\left(-2+ \eta_H  \right) \bar \mu^2 + \left.\partial_{\bar \rho} \,\,\overline{\text{Flow}}\left[	V_{\textrm{eff}}\right]\right|_{\bar \rho_0=0}\,,
\end{align}
where $\partial_t\equiv k \partial_k$, $\eta_H=-\partial_t\, Z_H/Z_H$ is the anomalous dimension of the Higgs field, and 
\begin{align}\label{eq:power-lawflowmu}
	\left.\partial_{\bar \rho} \,\,\overline{\text{Flow}}\left[	V_{\textrm{eff}}\right]\right|_{\bar \rho_0=0}=& \frac{1}{8\pi^2}\Bigg[ -\frac{\bar \lambda}{(1+\bar\mu^2)^2}\left(3-\frac{\eta_{\mathcal{G}^\pm}}{6}-\frac{\eta_{\mathcal{G}^0}}{12}-\frac{\eta_H}{4}\right)+ 3 y_\textrm{t}^2 \left(1-\frac{\eta_\textrm{t}}{5}\right)\notag\\[1ex]
	&\hspace{2cm}- \frac{18\, g_1^2}{80}\left(1-\frac{\eta_{Z^0}}{6}\right)- \frac{18\, g_2^2}{16}\left(1-\frac{\eta_{Z^0}}{18}-\frac{\eta_{W^\pm}}{9}\right)\Bigg]\,,
\end{align}
is the first $\bar{\rho}$-derivative of the right-hand side of the flow equation~\cite{Wetterich:1992yh} for the Higgs potential in a dimensionless form, $\overline{\text{Flow}}\left[V_{\textrm{eff}}\right]$, which has been derived in several works, see e.g.~\cite{Goertz:2023pvn,Pastor-Gutierrez:2022nki,Gies:2023jzd,Pawlowski:2018ixd}. Similarly, this second contribution can also be extracted from the diagramatic flow of the Higgs two-point function evaluated at vanishing external momenta, which encodes the relevant corrections to the mass. Here, for simplicity, we have dropped all fermionic contributions to the flow of the potential except the dominant one of the top quark. Furthermore, the $\eta_i$ are the anomalous dimensions entering from the regulated lines of the diagrammatic flows, see e.g.~\cite{Goertz:2023pvn,Pastor-Gutierrez:2022nki,Gies:2023jzd} for explicit computations in this context.

In \eqref{eq:flowmu}, we can distinguish two terms. The first is linear in $\bar \mu^2$ and contains the canonical dimension of the mass operator and the anomalous scaling of the field. The second term is not linear in the curvature mass and leads to a power-law scaling in the RG-flow. While here we discuss the dimensionless flow, this latter term appears proportional to the cut-off scale $k^2$ in the flow of the dimensionful analogue. Furthermore, this term is commonly referred to as the \textit{quadratic divergence} and its relevance deserves a detailed discussion provided in \Cref{sec:naturalness}. As a side note, the presence of the Higgs curvature mass as a threshold function in \eqref{eq:power-lawflowmu} makes explicit the mass-sensitive nature of the fRG which appears particularly beneficial in the description of near-threshold physics. In the broken phase, the Yukawa and gauge corrections are accompanied by the respective threshold functions which, after the Higgs mechanism turns effective, will naturally and progresively decouple the respective contributions.
 
Furthermore, the flow of the Higgs quartic coupling can also be obtained from the flow of the effective potential by performing an additional field derivative on the right-hand side of \eqref{eq:flowmu}, see~\cite{Pastor-Gutierrez:2022nki,Goertz:2023pvn} for its derivation. This reproduces the one-loop universal form and includes the threshold effects of the relevant off-shell modes.

\begin{figure*}[t!]
	\centering
	\includegraphics[width=1.\textwidth]{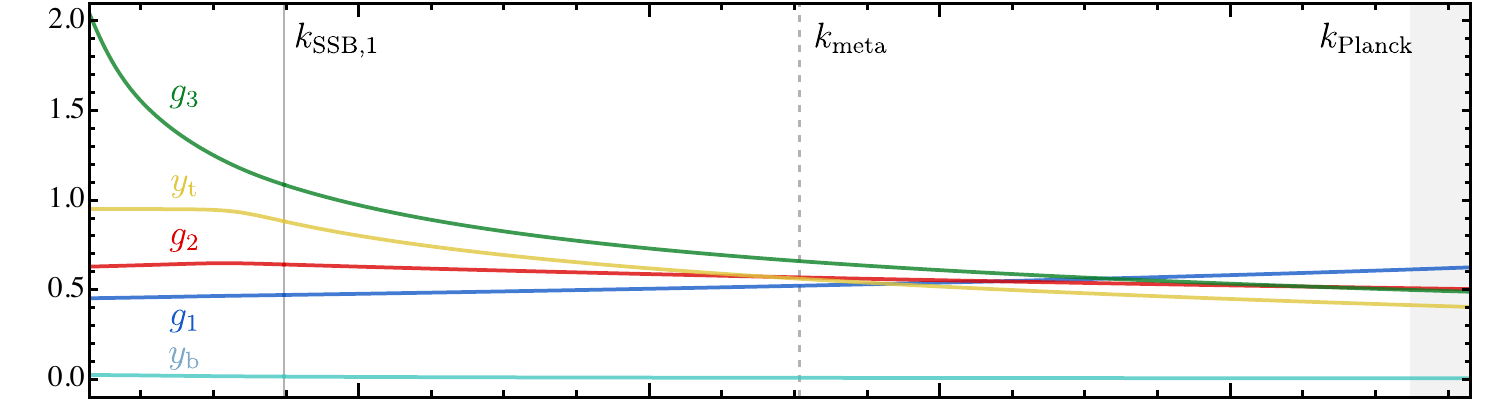}
	\includegraphics[width=1.\textwidth]{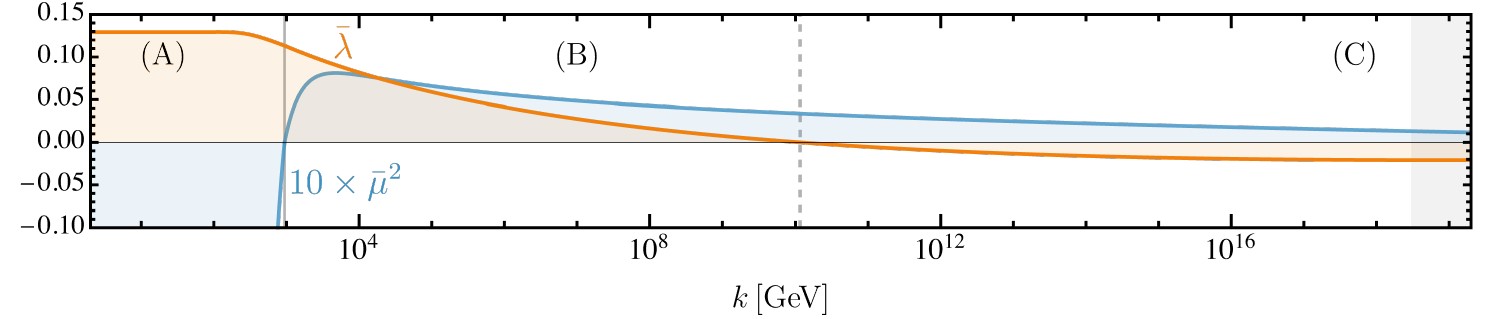}
	\caption{In the top panel, we show the trajectories of the SM gauge ($g_3,\,g_2,\,g_1$), top ($y_{\rm t}$) and bottom ($y_{\rm b}$) Yukawa couplings. These span from the Planck scale ($k_\textrm{Planck}\approx 10^{19}$\,GeV) down to below the dynamical chiral symmetry breaking scale ($k_{\chi\textrm{SB}}\approx 0.1$\,GeV), over the EW ($k_\textrm{SSB,1}\approx 10^{3}$\,GeV) scale. In the bottom panel, we display the trajectories of the dimensionless Higgs curvature mass ($\bar{\mu}^2$ in blue) and the quartic coupling ($\bar \lambda$ in orange). These two parameters define, in the present truncation, the shape of the average potential and hence the effective phases of the Higgs field. Three different regions are marked by plain ($\bar \mu^2=0$) and dashed ($\bar \lambda=0$) vertical lines, depending on the change of the small or large field curvature.}
	\label{fig:SM trajectories}
\end{figure*}

After this brief discussion on the derivation of the flows of the Higgs parameters within the SM, we proceed by showing in \Cref{fig:SM trajectories} the integrated flow trajectories which reproduce the experimental values measured below the TeV scale. In the top panel, we show the strong ($g_3$), weak ($g_2$) and hypercharge ($g_1=\sqrt{5/3}\, g_Y$) gauge couplings as well as the top ($y_\textrm{t}$) and bottom ($y_\textrm{b}$) Yukawa couplings. In the lower panel, we show the integrated flows for the Higgs quartic coupling ($\bar \lambda$) and curvature mass ($\bar \mu^2$). The flows span energy scales ranging from the Planck scale down to below the deep IR QCD scale ($k_{\chi\textrm{SB}}\simeq 0.1$\,GeV). For details on the truncation employed in the computation and the scale setting procedure followed here we refer to~\cite{Pastor-Gutierrez:2022nki}. 

In \Cref{fig:SM trajectories} we can identify three different regimes determined by the shape of the effective average potential, namely the effective potential at finite $k$. While the full effective potential is recovered as $k\to0$, the effective average Higgs potential displays the \textit{effective} shape at the respective scale and encodes the quantum fluctuations from $\LUV$ to $k$. We extract its shape from the integrated values of $\bar{\mu}^2$ and $\bar \lambda$ at a given scale. Starting at the Planck scale and flowing down, the curvature mass is positive and the Higgs self-coupling negative, see regime (C). This renders an effective shape of the Higgs potential with a metastable minimum where the Higgs field is found and expected not to tunnel. Moreover, at $k_\textrm{meta}\simeq10^{10}\,$GeV, $\bar{\lambda}$ turns positive and the effective shape of the Higgs potential recovers a trivially stable shape (regime (B) in \Cref{fig:SM trajectories}). At $k\simeq5\cdot 10^3$\,GeV the flow of the curvature mass rapidly changes signs and induces, at $k_\textrm{SSB,1}\simeq 950$\,GeV, a non-trivial minimum of the potential (see regime (A)). This corresponds to the Higgs' vacuum expectation value whose dimensionful form freezes at $v_{k\lesssim 200\textrm{GeV}}=246\,$\,GeV. This non-trivial vacuum generates the masses of the EW gauge bosons and fermions via the Higgs mechanism~\cite{Englert:1964et,Higgs:1964pj,Guralnik:1964eu}. For the cutoff profile of $v$ and all the SM Euclidean masses, see~\cite{Pastor-Gutierrez:2022nki}. 

As the sliding cutoff scale decreases below $k<k_\textrm{SSB,1}$, we find a continuous and progressive flattening of the RG running of the couplings given by the smooth decoupling of degrees of freedom, with threshold scales $m_{i}\propto g_i v$. For example, the top Yukawa coupling $y_\textrm{t}$ freezes in at $k\simeq 165$\, GeV which corresponds to the top quark Euclidean mass in the fRG scheme~\cite{Pastor-Gutierrez:2022nki} and matches the physical pole mass~\cite{ParticleDataGroup:2024cfk}.  We note here that in the lower pannel of \Cref{fig:SM trajectories} we display the dimensionless curvature mass instead of the dimensionful one (see \Cref{fig:SM phase diagram mHiggs no EWSB}). The rapid growth towards the IR is explained given that the dimensionful mass is constant in the $k\to 0$ limit, hence $\bar{\mu}^2_{k\to 0}\propto k^{-2}$. 

With \Cref{fig:SM trajectories} at hand we would like to emphasize that the fRG flow and in particular the treatment of the curvature Higgs mass $\bar \mu^2$, provides several beneficial features that carry interesting physical information. In contrast to the common approach used, the curvature mass provides information about the effective shape of the Higgs potential at different energy scales. Large part of this information is encoded in the power-law component of the flow of the curvature mass in \eqref{eq:power-lawflowmu}. An example of this is the zero-crossing of the small-field curvature which is only possible due to the power-law divergence in the associated flow. Strictly speaking, this evolution does not represent a physical phase transition since $k$ is not a physical parameter, but it mimics it. In the theory of critical phenomena, such transitions occur when a physical parameter such as the temperature of a system is varied and can be quantified via the change of an order parameter, which in the present context can be taken to be the Higgs vacuum expectation value. 

The fRG regulator, which is quadratic in the fields, acts as a mass-like IR sliding cutoff, progressively incorporating quantum corrections as $k\to 0$. This regularization procedure is indeed analogous to the effect of temperature on thermal fluctuations which also naturally introduces an IR cutoff. More precisely, the role of thermal corrections as temperature decreases is analogous to the RG steps progressively incorporating quantum fluctuations from UV to IR. The arrangement of off-shell contributions along the RG flow mirrors their behavior under varying physical temperature. In essence, thermal corrections \textit{undo} the quantum effects observed at the IR cutoff level. To clarify this relation between temperature and the IR cutoff, consider the finite temperature regularized propagator of the Higgs field. It reads
\begin{align}\label{eq:regpropT}
G_{H,k}(T,p^2)=\frac{1}{Z_{H,k}}\,\frac{1}{\omega^2_n+{\vec p}\,^2\,(1+r_k(\vec p))+m_{H,k}^2}\,,
\end{align}
where $\omega_n=2 n \pi T$ are the respective Matsubara modes of the Higgs field, $m_{H,k}$ is the Euclidean mass and $r_k$ a dimensionless spatial regulator function. The integration of quantum fluctuations is first performed along the $k$ flow by means of $r_k$ and the respective flow is gapped at the threshold of the Higgs mass. For finite $T$, the effect of Matsubara modes in \eqref{eq:regpropT} introduces a similar gap, freezing in quantum fluctuations. In the SM Higgs potential, the leading quantum and thermal corrections stem from fermions, scalars, and EW bosons respectively, with the top quark contribution being the dominant one. Since the phase transition is tied to the scalar potential, it is reasonable to approximate the cutoff to bosonic thermal corrections. Using this identification, the relevant mode maps to the zero-temperature cutoff as $k \approx 2\pi T$. This approximately yields a critical temperature for the EW phase transition of
\begin{align}\label{eq:critialTEW}
	T^\textrm{crit}_\textrm{SSB,1} \sim k_\textrm{SSB,1}/(2\pi) \approx 150 \,\textrm{GeV}\,.
\end{align}
This rough estimate is surprisingly within $6\%$ agreement with lattice computations of non-Abelian-Higgs theories which find $T^\textrm{crit} = 159.6 \pm 0.1 \pm 1.5 \,\textrm{GeV}$~\cite{DOnofrio:2015gop}. The identification here discussed is not new but well known from various finite-temperature fRG computations, such as~\cite{Fister:2011uw,Fister:2011um}, which discuss the impact of thermal fluctuations on the effective average action, and~\cite{Fu:2019hdw,Braun:2005uj,Khan:2015puu,Reichert:2017puo,Hawashin:2024dpp,Tolosa-Simeon:2025fot}, which provide examples in different contexts. However, while the fRG cutoff qualitatively resembles temperature, this analogy should not be interpreted quantitatively. Determining the precise order of the phase transition and the exact critical temperature requires explicit thermal contributions.
Nevertheless, the approach allows to get access to the features of the phase structure in an economical and reasonably accurate way. 

\section{Phase diagram of the Standard Model class of theories}\label{sec:SMPT}
\begin{figure*}[t!]
	\centering
	\includegraphics[width=1.\textwidth]{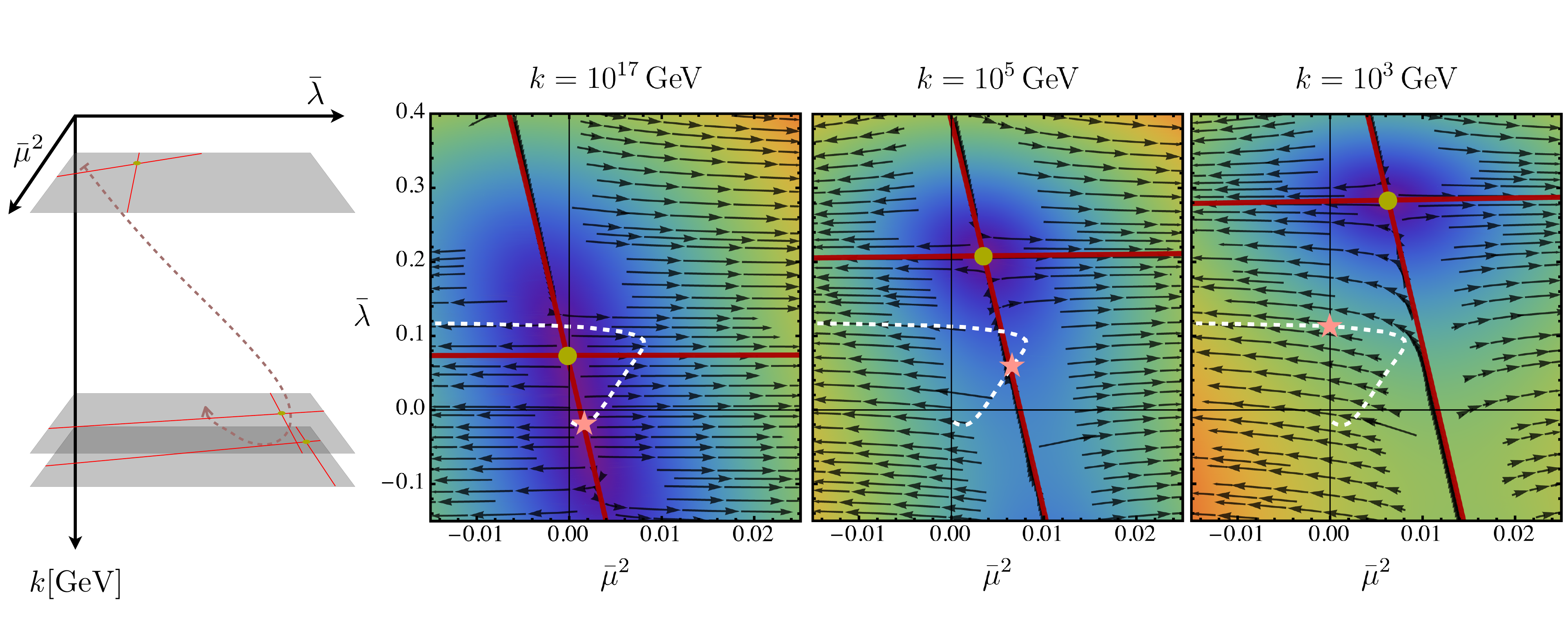}
	\caption{Phase diagram of the SM class of theories projected on the plane of scalar sector parameters ($\bar\mu^2$--$\bar\lambda$). The left-most cartoon depicts the flow in a sub-plain of theory space. On the plots, the black arrowed lines indicate the flows from UV to IR for the parameter space. The yellow dot and red lines correspond to the partial fixed-point solutions and the separatrix as defined in \eqref{eq:FPHiggs}, \eqref{eq:separatrix1} and \eqref{eq:separatrix2}, respectively. From left to right, we display slices of theory space at $k=10^3$, $10^{5}$ and $10^{17}$\,GeV. The pink star indicates the position of the SM trajectory at the respective scales and the white dashed line the full trajectory in the multi-dimensional volume projected on the plane of couplings. }
	\label{fig:SM phase diagram}
\end{figure*}
The SM is found within the class of gauge-fermion QFTs, with a priori nineteen relevant free parameters. To study this multi-variable class of theories, one may simplify the phase diagram analysis by projecting onto a sub-dimension. Following this idea, in \Cref{fig:SM phase diagram}, we depict the SM phase diagram projected onto the $(\bar \mu^2-\bar \lambda)$--plane of scalar parameters at different RG-times or energy scales: $k=10^3$, $10^{5}$, and $10^{17}$\,GeV, from left to right. The black arrowed lines give the instantaneous flows, from UV to IR, of SM-like theories, and are determined by the flow equations. We emphasize that the phase diagrams of \Cref{fig:SM phase diagram} correspond to snapshots (at fixed RG time) of a particular plane of theory space and these will change as we flow among different RG times, since other dimensions of the theory (e.g.~gauge and Yukawa couplings) also run and feed information back into the scalar couplings. The white dashed line denotes the integrated trajectory of the SM, projected onto the plane of scalar sector couplings.

This phase diagram allows us to identify different structures and phases. First, each of the quadrants corresponds to different shapes of the effective average potential. We have argued above how this resembles the effect of a physical temperature. 

Moreover, there is a fixed-point-like solution, depicted by a yellow point that satisfies
\begin{align}\label{eq:FPHiggs}
	&\partial_t \bar \lambda \left(\bar \mu^{*\,2},\bar\lambda^*,\vec g\right)=0&& \textrm{and}&&&\partial_t \bar \mu^2 \left(\bar \mu^{*\,2},\bar\lambda^*,\vec g\right)=0\,,
\end{align}
and has one attractive and one repulsive direction in this plane of couplings. Here $\vec{g}$ stands for the remaining set of marginal couplings, i.e., gauge and Yukawa. We stress that this is not a \textit{full} fixed-point solution of the theory, given that it moves along with the RG time and that $\vec g\neq \vec g^*$. Only two of the 19 parameters of the theory satisfy this condition. Consequently, we denote this as a \textit{partial fixed point of dimension 2}. Furthermore, in \Cref{fig:SM phase diagram}, there are further partial fixed-point solutions of dimension one, which appear as lines on the plane. These are depicted by red lines and satisfy
\begin{align}\label{eq:separatrix1}
	&\partial_t \bar \lambda \left(\bar \mu^{2},\bar\lambda,\vec g\right)\neq0&& \textrm{and}&&&\partial_t \bar \mu^2 \left(\bar \mu^{2},\bar\lambda,\vec g\right)=0\,,
\end{align}
for the vertical one and 
\begin{align}\label{eq:separatrix2}
	&\partial_t \bar \lambda \left(\bar \mu^{2},\bar\lambda,\vec g\right)=0&& \textrm{and}&&&\partial_t \bar \mu^2 \left(\bar \mu^{2},\bar\lambda,\vec g\right)\neq0\,,
\end{align}
for the horizontal one. The described structures manifest important features of the theory, which we will study one by one in the following Sections. In \Cref{fig:SM phase diagram}, the partial fixed point has an IR attractive and an IR repulsive direction. The former is provided by the large canonical scaling of the curvature mass and the latter by the Higgs quartic coupling. For the moment, we do not comment on the IR/UV repulsiveness in the other directions of theory space. 

One can identify a boundary trajectory that separates theories with fundamentally different IR scenarios: those with a non-trivially stable shape of the Higgs potential ($\bar \rho_{0,\,k\to0}\neq0$) and others with a trivial one ($\bar \rho_{0,\,k\to0}= 0$). The special trajectory separating these two sets of solutions is usually denoted as the \textit{separatrix}. Furthermore, note that in the projection shown above it coincides with the partial fixed point solution of \eqref{eq:separatrix1}. However, generally speaking, such an agreement between separatrices and partial fixed points holds only in the neighborhood of the fixed point.

From the partial fixed-point condition for the curvature mass, it follows that 
\begin{align}\label{eq:partialFPmu}
\partial_t \bar \mu^2 =0 \qquad\forall k&&\Rightarrow&&&  \bar \mu^2 \approx \frac{\left.\partial_{\bar \rho} \,\,\overline{\text{Flow}}\left[	V_{\textrm{eff}}\right]\right|_{\bar \rho_0=0}}{\left(2- \eta_H  \right)}\,,
\end{align}
which renders the cut-off evolution of the mass parameter as a function of the other couplings in the theory. While in \eqref{eq:partialFPmu} the relation is only approximate, given that we have neglected the $\bar{\mu}^2$ dependence entering the diagrammatic flows (e.g.  via the subdominant threshold functions), these can be accounted for leading to a resummed non-perturbative form. Setting $\vec{g}=0$ in \eqref{eq:partialFPmu}, we recover the pure scalar theory in four dimensions where the fixed point in \Cref{fig:SM phase diagram} turns to the Gau\ss ian one of the free theory, see~\cite{Yamada:2020bqe,Aoki:2012xs} for such an analysis. Furthermore, the cutoff profile of the curvature mass is now given, on this trajectory, as a function of the other couplings, $\bar \mu^2(\bar \lambda,\vec{g})$. While the partial fixed point condition is rather straightforward and carries relevant information that will become apparent later on, the question of whether trajectories in theory space satisfy such a condition for all scales is more subtle. Although this can be found to be the case for simple theories, such as those of pure scalar and Yukawa systems, for more elaborated actions like in the SM, determining the fulfillment of such a condition becomes a challenging task. 

The partial fixed point condition in \eqref{eq:partialFPmu} is also commonly known as the \textit{massless trajectory} given that on this the scalar field never decouples, neither with a positive nor a negative curvature mass. Note that while in the pure scalar case the mass parameter $\bar \mu^2_{k\to0}\to\bar \mu^{*\,2} =0$, if (3.4) is fulfilled, in SM-like systems $\bar \mu^2_{k\to0}\to \bar \mu^{*\,2}\neq 0$. Nevertheless, in both cases a non-decoupling of the Higgs scalar is realized, and no explicit scale is generated. We may refer to this particular solution as the trajectory \textit{on} the critical surface~\cite{Aoki:2012xs,Wetterich:2019qzx} and it minimizes the impact of the relevant scaling.
Although partial fixed points such as the one in \eqref{eq:partialFPmu}, do not have to be related to symmetries of the classical action, they usually are. In the present case with the curvature mass, it is clear how the massless condition renders scale symmetry~\cite{Wetterich:1983bi,Wetterich:1990an,Wetterich:2019qzx}.

\begin{figure*}[t!]
	\centering
	\includegraphics[width=.6\textwidth]{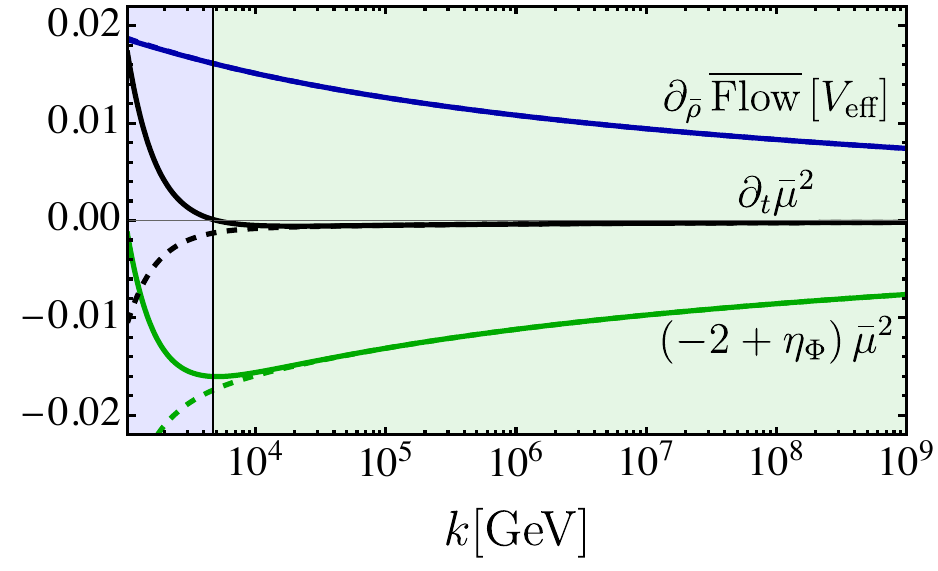}
	\caption{Total ($\partial_{t} \bar \mu^2$, plain black line) and partial (plain blue and green lines) contributions of the flow of the Higgs curvature mass for the SM trajectory. The power-law contribution or critical surface is depicted in blue ($\left.\partial_{\bar \rho} \,\,\overline{\text{Flow}}\left[	V_{\textrm{eff}}\right]\right|_{\bar \rho_0=0}$), and the logarithmic ($(-2+\eta_H)\bar \mu^2$) in green. The symmetric (shaded green) and broken (shaded blue) regimes are marked, where logarithmic and power scalings respectively dominate. The dashed lines show a neighbouring solution to the SM one which leads to no EWSB, as discussed in \Cref{fig:SM phase diagram mHiggs no EWSB}. }
	\label{fig:flowmu}
\end{figure*}

Returning to the SM phase diagram, in \Cref{fig:SM phase diagram} we observe that all trajectories emanate from the critical surface as they flow from the UV to the IR. For the SM trajectory initiated at the Planck scale, the curvature mass is initially close to the separatrix and its flow exhibits a logarithmic growth towards the IR. At scales $k\sim5\cdot 10^3$\,GeV, the logarithmic contribution is surpassed by the power law one with opposite sign. The rapid departure away from the critical surface is triggered by the large canonical dimension of the Higgs bilinear and its competition with the quadratic part of the flow, see \eqref{eq:flowmu}. This behavior, depicted in \Cref{fig:flowmu}, leads to the zero crossing of the curvature mass and therefore to the transition between the symmetric and broken phases of the effective average potential. 

Moreover, the power-law scaling of the flow encoded in $\partial_{\bar \rho}\, \overline{\text{Flow}}\left[V_{\text{eff}}\right]$ is rather independent of the Higgs curvature mass and is instead almost fully determined by the Yukawa and gauge couplings, as shown in \eqref{eq:power-lawflowmu}. Furthermore, this term crucially determines the position of the critical surface, which sets a reference value of $\bar \mu^2_{\LUV}$ that one must approximate more and more to achieve an increasingly precise matching between the logarithmic and power-law scalings which renders the desired separation of scales. Therefore, a theory with a large-scale separation between $\LUV$ and $\kSSBone$ will live, in the UV, very close to the critical surface. On the contrary, theories with UV values far from the critical surface quickly develop a mass scale, either via a hard mass or a non-trivial $v$, see \Cref{fig:SM phase diagram mHiggs no EWSB}. This discussion illuminates the tight relation between the hierarchy problem and the near-quantum criticality of SM-like theories, which we will elaborate on in \Cref{sec:quantumPT}.  

Power-law corrections in dimensionful parameters are treated differently across renormalization schemes. In mass-independent schemes, such as dimensional regularization, these corrections do not appear explicitly in the RG equations. Instead, they arise at physical scales as a result of integrating out heavy modes. When matching effective descriptions above and below a threshold scale, power-law corrections—which are absent in the flow—enter as threshold contributions. This matching, accounting for relevant corrections, is equivalent to specifying the position of the theory relative to the critical surface. In contrast, mass-dependent schemes such as the fRG, retain power-law corrections along the RG flow. This allows the flow to dynamically track both the position of the critical surface and the distance of the target theory from it, without requiring a specific UV completion. Importantly, while the precise value of the power-law corrections—and thus the location of the critical surface—is scheme dependent, the distance of a theory to the critical surface (or to any other theory) is determined solely by logarithmic corrections and is scheme independent~\cite{Aoki:2012xs,Wetterich:2019qzx,Yamada:2020bqe}. In this sense, the critical surface defines a reference frame with respect to which physical observables remain invariant.

In summary, the specification of the critical surface requires technical tuning, which may vary across approaches. However, physical tuning—captured by the distance to the critical surface—is scheme independent and thus physically meaningful. These notions of technical and physical tuning have been thoroughly discussed in \cite{Aoki:2012xs} in the context of Wilsonian schemes and also at a more conceptual level in \cite{Wells:2025hur}. For this reason, in the following we will discuss the sensitivity of IR observables with respect to relative distances. 

\section{Quantum phase transition and criticality}\label{sec:quantumPT}
We continue the analysis of the phase diagram of the SM class of theories by investigating the change of IR physics when dialing over neighboring theories. We begin by shifting the Higgs curvature mass with respect to the value that reproduces the physical SM trajectory ($\bar\mu^2_{k,\,\textrm{SM}}$) at a UV scale $\LUV$. In other words, we consider relative variations,
\begin{align}\label{eq:relativemu0}
	\Delta \bar \mu^2_\textrm{rel} (\LUV)=\left.\frac{\bar \mu_k^2-\bar \mu^2_{k,\,\textrm{SM}}}{\bar \mu^2_{k,\,\textrm{SM}}}\right|_{k=\LUV} \,,
\end{align}
and use this parameter as our control parameter for theory selection~\cite{Gies:2023jzd}. Strictly speaking, this exercise is equivalent to dialing over different theories and reflects the relation between scales in a certain QFT. However, as will be done in \Cref{sec:naturalness}, the UV cutoff can be upgraded to carry a physical meaning, for example that of a threshold where a heavy particle decouples. Consequently, the exercise of dialing boundary conditions at $\LUV$ is equivalent to considering different embeddings into a more complete theory that includes new physics. 

\begin{figure*}[ht!]
	\centering
	\includegraphics[width=.32\textwidth]{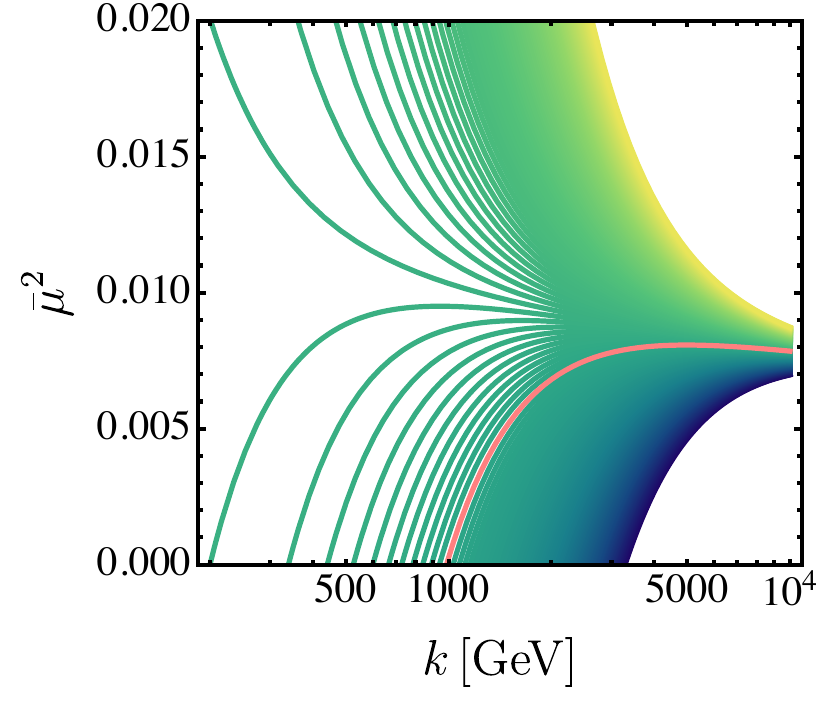}
	\includegraphics[width=.32\textwidth]{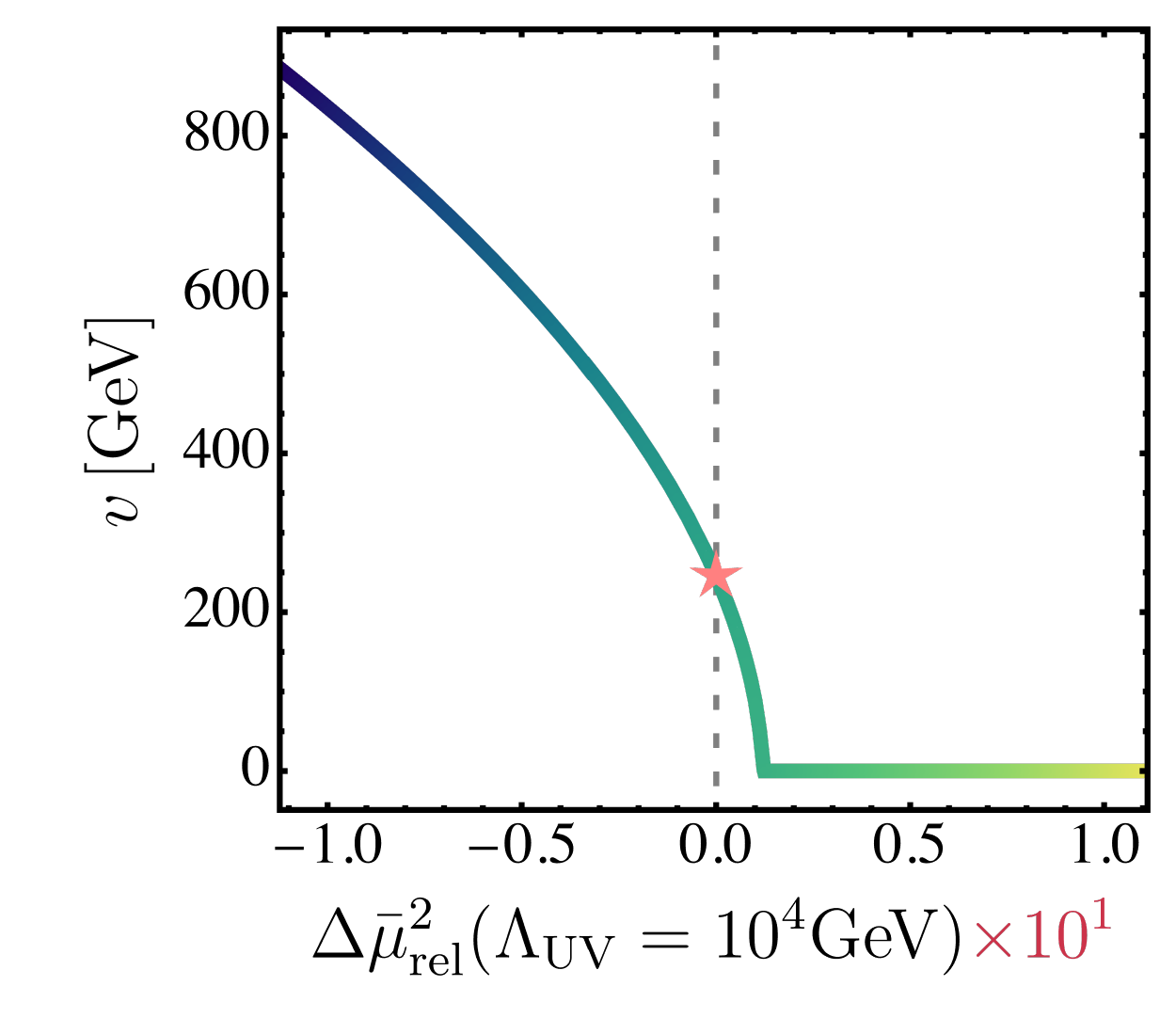}
	\includegraphics[width=.32\textwidth]{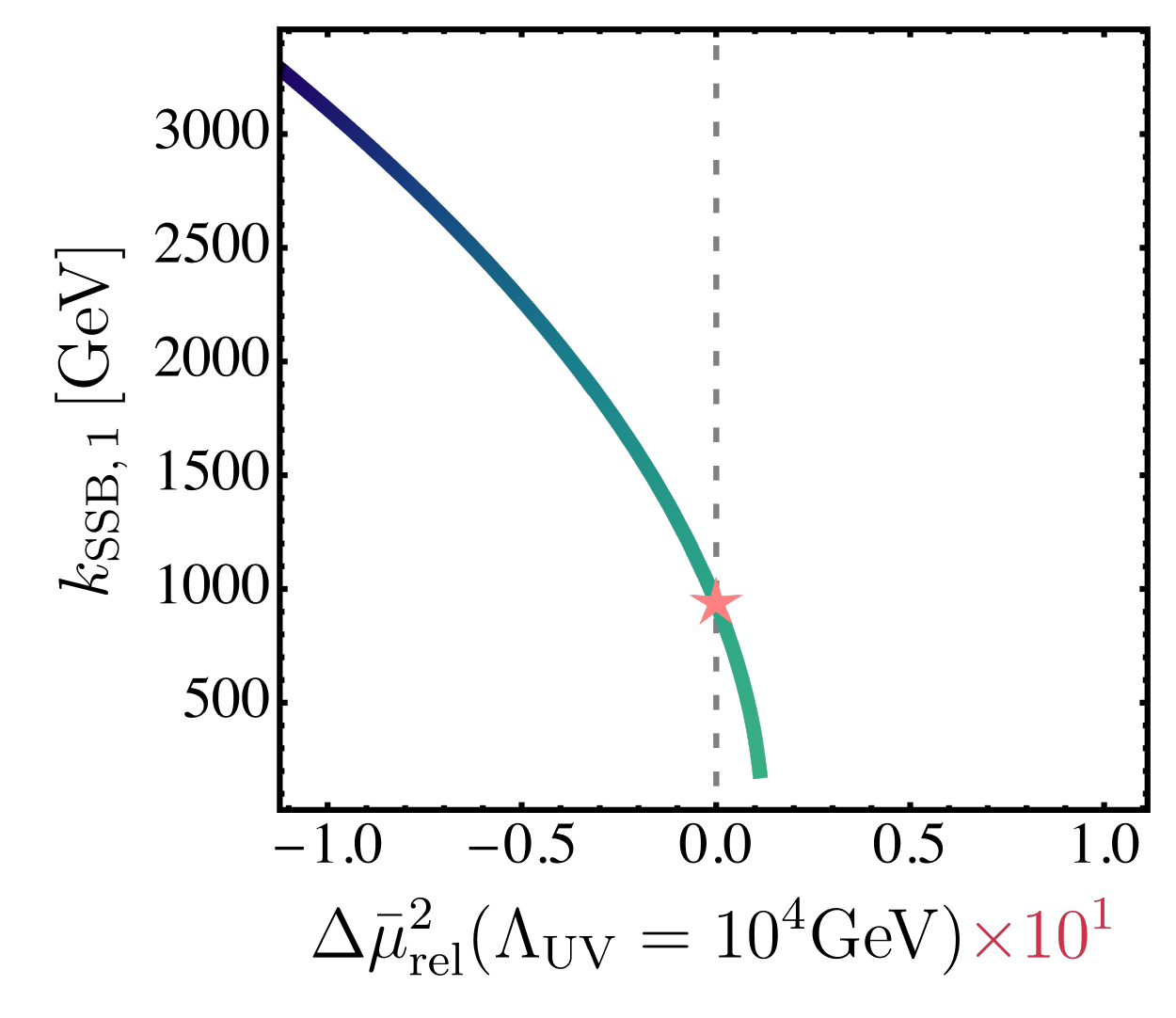}
	\includegraphics[width=.32\textwidth]{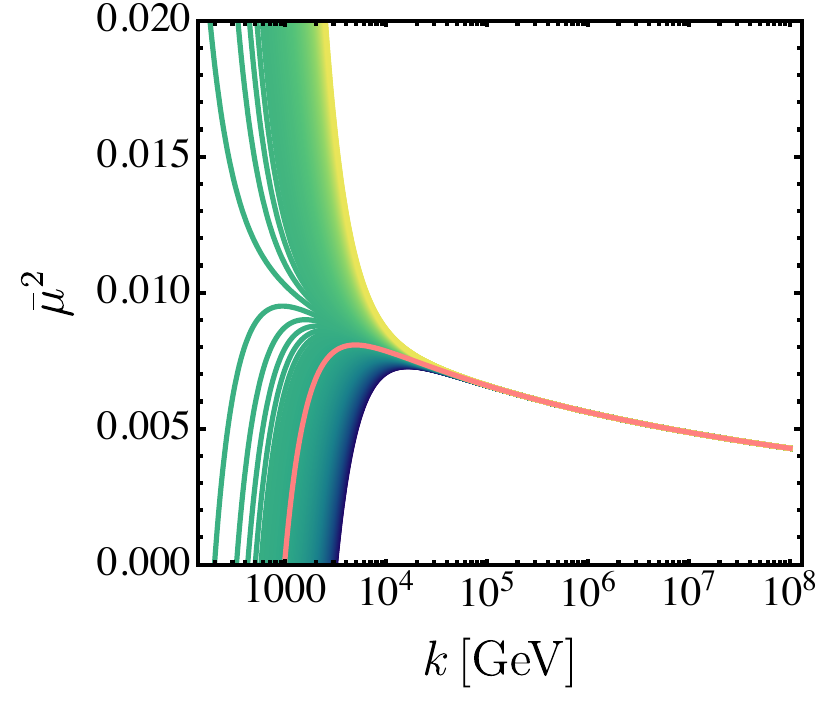}
	\includegraphics[width=.32\textwidth]{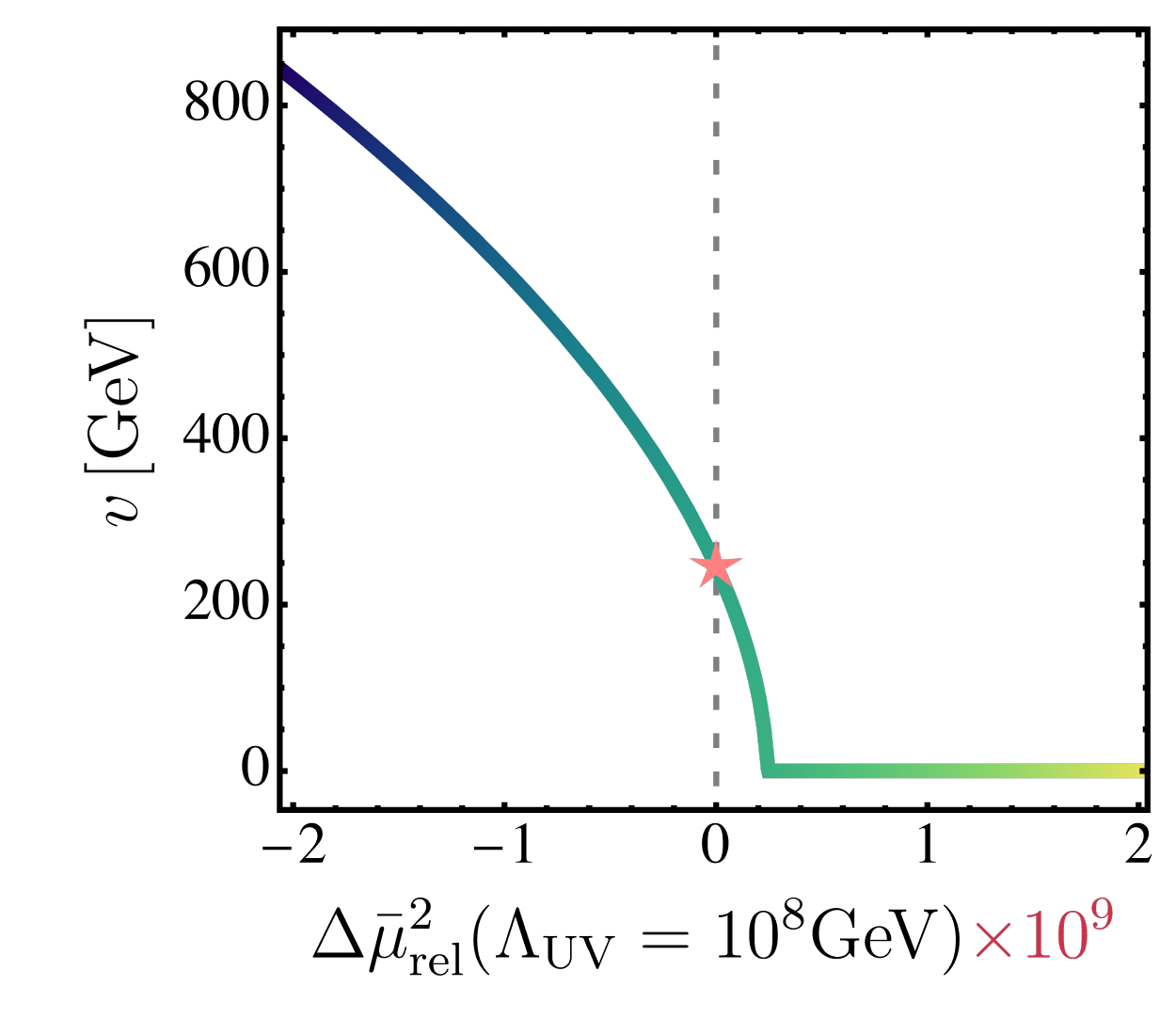}
	\includegraphics[width=.32\textwidth]{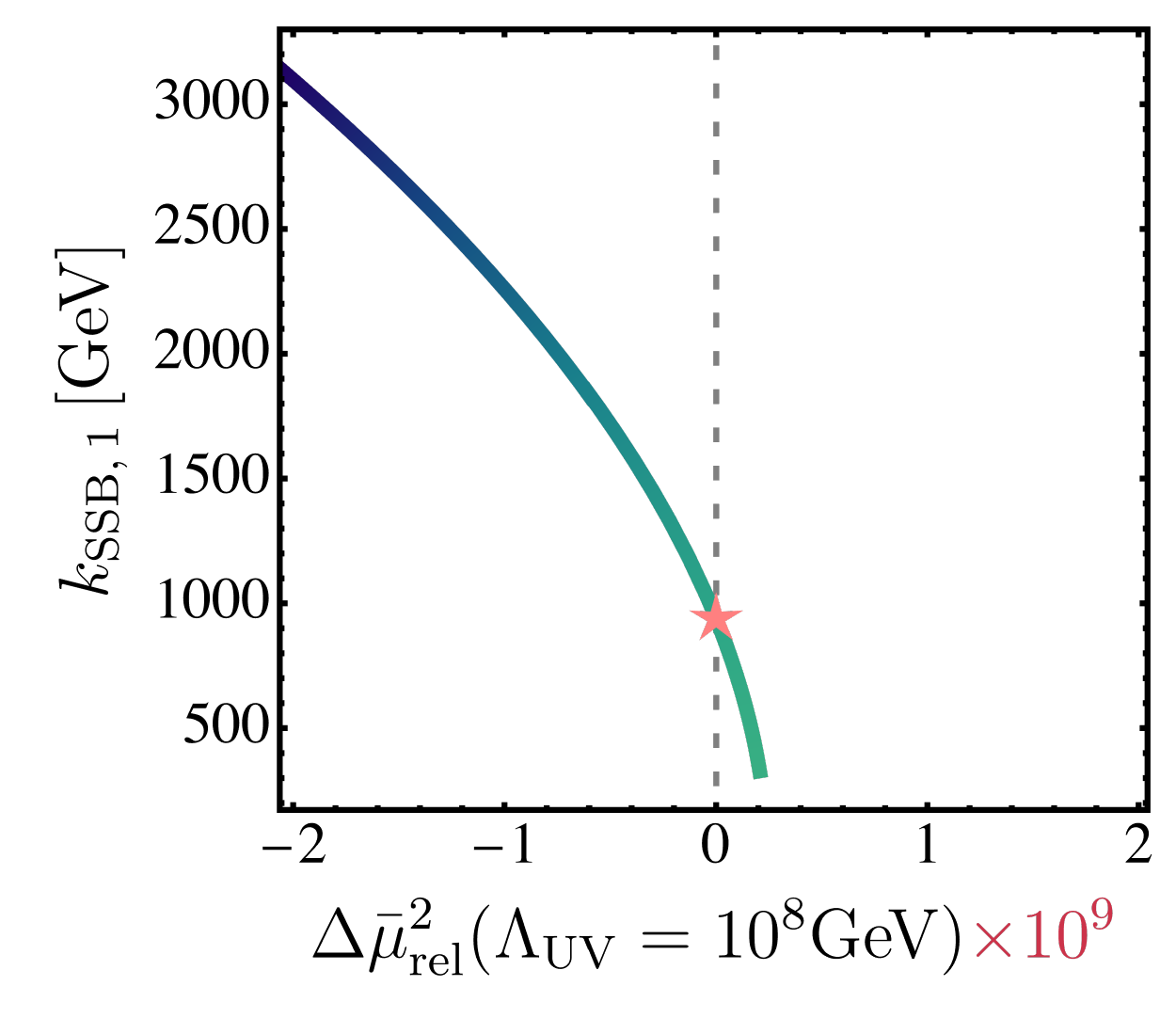}
	\includegraphics[width=.32\textwidth]{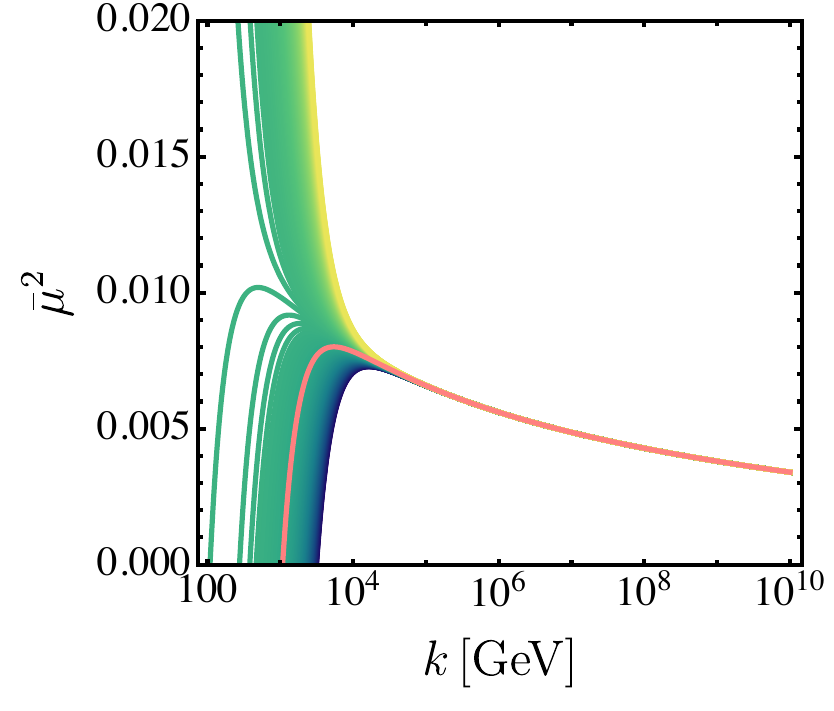}
	\includegraphics[width=.32\textwidth]{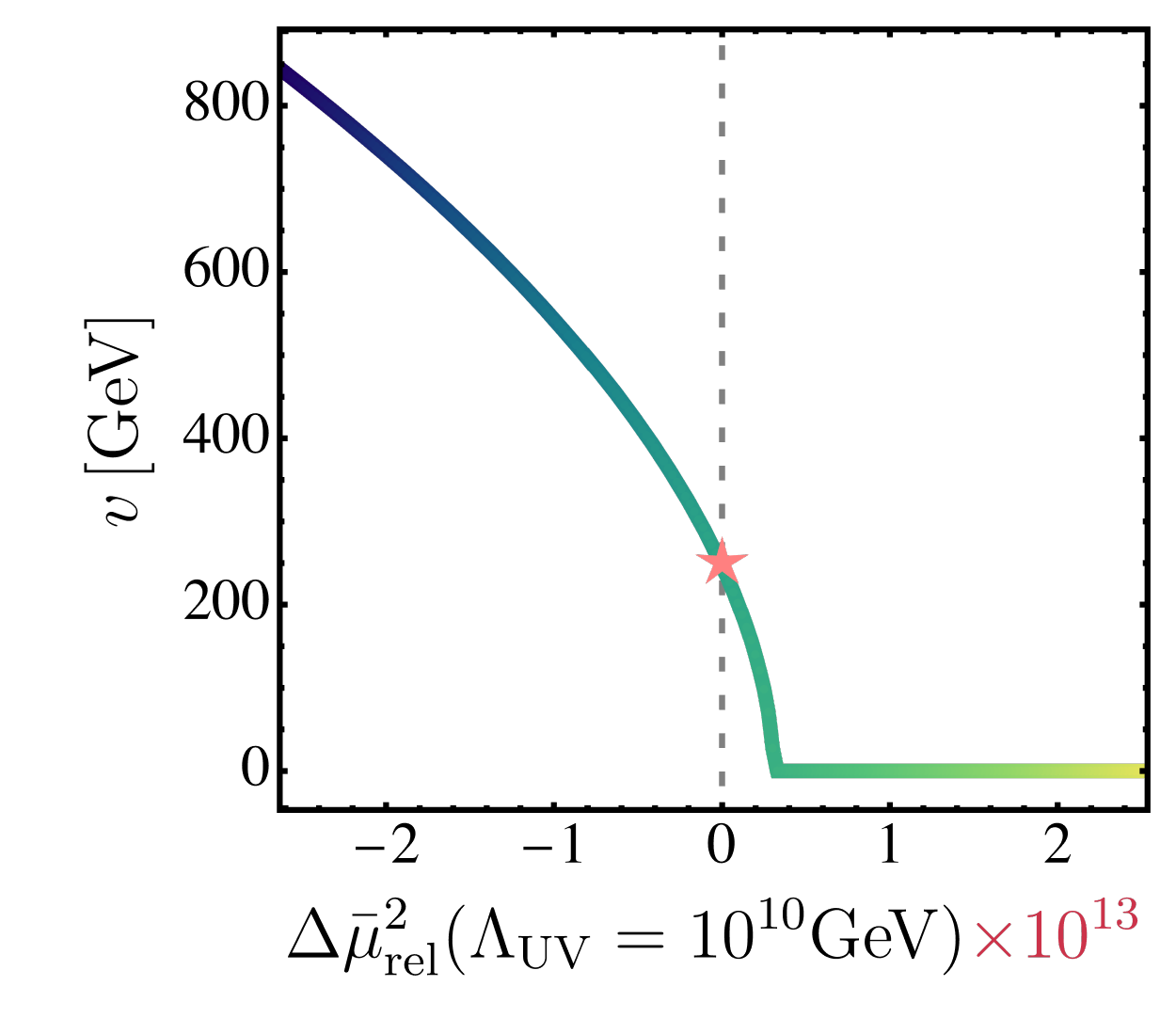}
	\includegraphics[width=.32\textwidth]{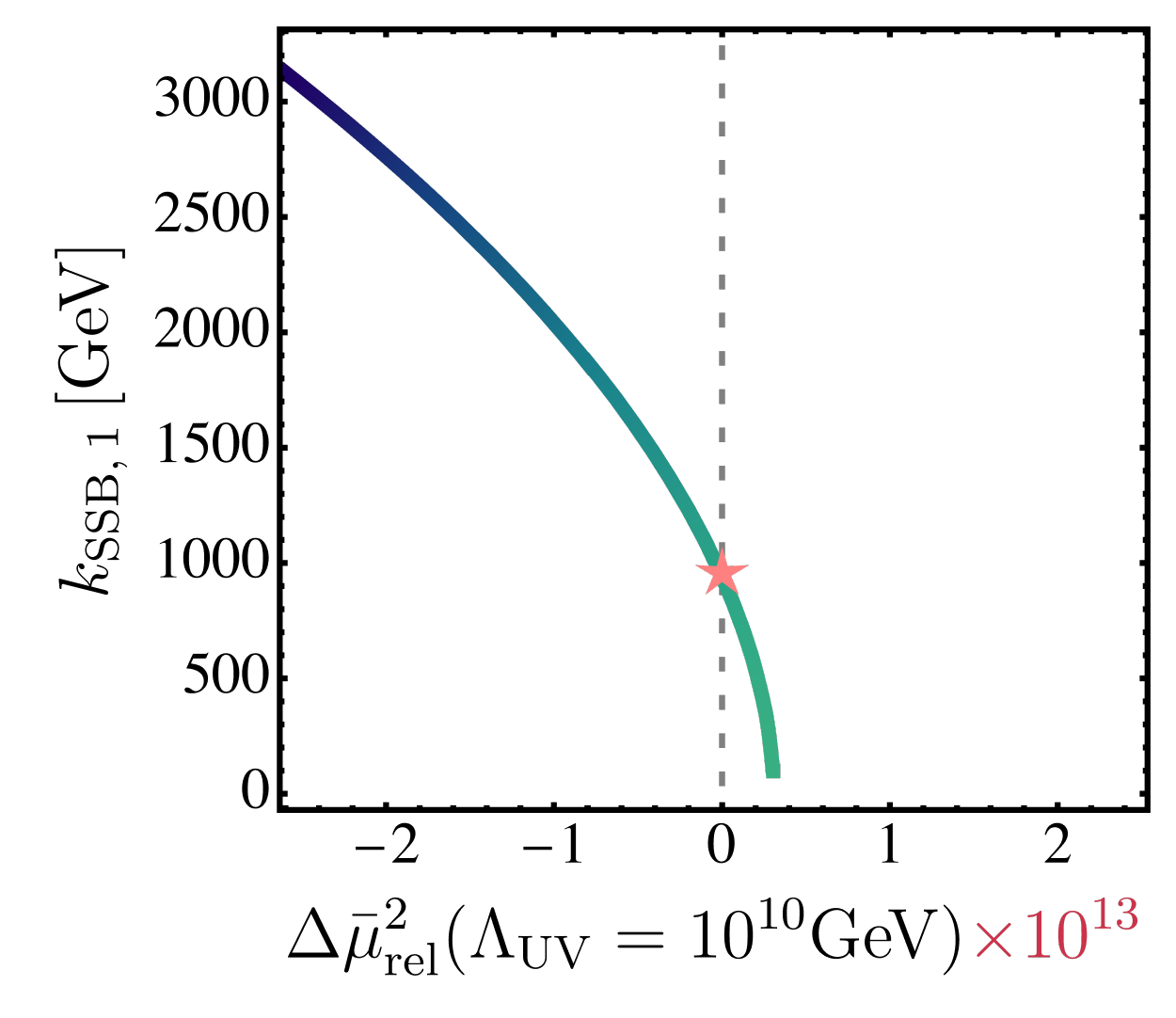}
	\caption{In the left-most panels, the Higgs curvature mass trajectories for various boundary conditions (yellow to blue lines) around the physical one (pink lines and stars) are shown. In the middle (rightmost) panel, we show the Higgs vacuum expectation value $v$ at $k\to0$ (value of the $\kSSBone$ scale), as a function of the relative variation of the boundary condition with respect to the physical trajectory ($\Delta \bar \mu^2_\textrm{rel} (\LUV)$). From top to bottom, each row corresponds to a different choice of  $\LUV=\{10^4,10^8,10^{10}\}$\,GeV, respectively. See text for details. 
	}
	\label{fig:finetuning SM}
\end{figure*}

In the left-most column of \Cref{fig:finetuning SM}, we depict the integrated flows of the Higgs curvature mass for different relative deviations from the SM and starting the flow at different $\LUV$ scales (shown in different rows). In the middle and right columns, the Higgs vacuum expectation value and the $\kSSBone$ scale are shown as functions of the relative variation \eqref{eq:relativemu0}, respectively.  These plots show that an exceptionally precise tuning of order ${\cal O}(\kSSBone^2/\LUV^2)$ needs to be performed at the level of UV boundary conditions in order to fix realistic IR parameters.  This is nothing but a manifestation of the large sensitivity of the IR physics on the UV sourcing the hierarchy problem. Quantifying this issue is within the scope of the present work and will be explored in \Cref{sec:naturalness,sec:newPhysicsdeformations}.

Furthermore, two distinct IR regimes can be realized depending on the values of the UV boundary conditions, one with a broken phase at $k\to0$ and another that remains in the symmetric regime. As described in the previous section, the boundary trajectory separating the two IR phases corresponds to the critical surface, which flows into the IR partial fixed point at $k\to0$. We denote the boundary value of this specific trajectory at $\LUV$ as $\bar{\mu}^2_{\LUV,\,\textrm{crit}}$. In the middle and right panels of \Cref{fig:finetuning SM}, this corresponds to the left-most point with $v\simeq0$. For $\bar \mu^2_{\LUV} > \bar{\mu}^2_{\LUV,\,\textrm{crit}}$, the Higgs potential remains in the symmetric phase and no Higgs mechanism takes place. In contrast, for $\bar\mu^2_{\LUV} < \bar{\mu}^2_{\LUV,\,\textrm{crit}}$, trajectories enter the broken phase. As $\bar\mu^2_{\LUV}$ decreases, $\kSSBone$ and $v$ increase relative to the EW scale, effectively bringing $\LUV$ and $\kSSBone$ closer together. 

The existence of distinct phases in the class of theories at $k \to 0$ signals the presence of a \textit{quantum phase transition}. For a review, see~\cite{Sachdev:2011fcc}, and for examples in various contexts studied with the fRG, see e.g.~\cite{Wetterich:2007ba,Sondenheimer:2016vko,Gies:2023jzd,Hawashin:2024dpp,Gies:2009az}. It is essential to differentiate between quantum and physical phase transitions; the latter occurs for a specific theory as a physical variable, such as temperature, pressure, or chemical potential, is varied. Quantum phase transitions share analogous properties and can be described using the theory of critical phenomena~\cite{ZinnJustin:2002ru}. In this framework, the Higgs vacuum expectation value serves as the order parameter, while the parameters or couplings of the effective action at a UV scale play the role of the physical variables (i.e. temperature, pressure, etc.). Note that in \Cref{fig:finetuning SM}, we considered only one control parameter, namely the curvature mass at a high-energy scale, but others shall be used as well, as will be done in \Cref{sec:naturalness}.

The scaling of the vacuum expectation value near the critical trajectory can be universally described by the theory of critical phenomena and the scaling relations give~\cite{Gies:2023jzd,ZinnJustin:2002ru}
\begin{align}\label{eq:QPTscaling}
	v \propto \left|\Delta \bar \mu^2_{\textrm{crit}}(\LUV) \times \LUV^2 \right|^\beta\,,
\end{align}
where $\beta$ is the critical exponent of the quantum phase transition, and $\Delta \bar \mu^2_{\textrm{crit}}$ is the control parameter, defined as the distance to the critical surface,
\begin{align}\label{eq:Deltasep}
	\Delta \bar \mu^2_{\textrm{crit}} = \bar \mu_{\LUV}^2 - \bar \mu_{\LUV,\,\textrm{crit}}^2\,,
\end{align}
analogously to the previously defined $\Delta \bar{\mu}^2_{\rm rel}(\LUV)$. The critical exponent $\beta$ is related to the properties of the scalar field at the critical scale~\cite{Gies:2023jzd}, namely at $\kSSBone$ where $\bar \mu^2_{\kSSBone} = 0$ and the correlation length is infinite.  The quantum phase transition in the SM class of theories appears to be of second-order, as seen in \Cref{fig:finetuning SM} and studied in~\cite{Gies:2023jzd}. The nature of this transition is governed by the partial IR fixed point (yellow point in \Cref{fig:SM phase diagram}). Depending on its position in theory space, the transition could, in principle, take a different order (e.g., first-order). For the trajectory reproducing the SM in the IR, $\left.\eta_H\right|_{k_{\textrm{SSB},1}} \approx 0.017$, which is close to the free theory, yielding a critical exponent of $\beta \approx 1/2$.

In obtaining the results in \Cref{fig:finetuning SM}, we stopped the flow at cutoff scales of $k\simeq10$\,GeV given that the strong QCD dynamics are highly dependent on the current quark masses sourced by the Higgs mechanism.
To properly understand the IR limit of this near-chiral SM, it is necessary to carefully assess the intricate interplay of the Higgs, EW and chiral bound states in the strongly coupled regime of chiral QCD theories~\cite{Goertz:2024dnz}. While resolving the full, dynamical and strong system of RG flow equations is within the scope of the present framework, this has recently been the topic of study in~\cite{Gies:2023jzd}. There, the dynamics induced by the Higgs boson and the driven by the bound state formation of quarks, have been studied on the same footing and, for the first time, from first principles. Understanding such interplay is highly relevant to decrypt the properties of the quantum phase transition such as its order. Moreover, the results of~\cite{Gies:2023jzd} show indication of a cross-over EW phase transition which is so sharp that it is practically indistinguishable from a second-order one. This is in qualitative agreement with the results shown in \Cref{fig:finetuning SM}.  Additionally, the authors showed that the fine-tuning is significantly relaxed as the Higss EWSB scale is drawn towards the strong dynamics scale. This is driven by the increase in the Higgs' anomalous dimension which effectively distances the critical surface by reducing the total dimensionality of the mass operator. 

\begin{figure*}[t!]
	\centering
	\includegraphics[width=.42\textwidth]{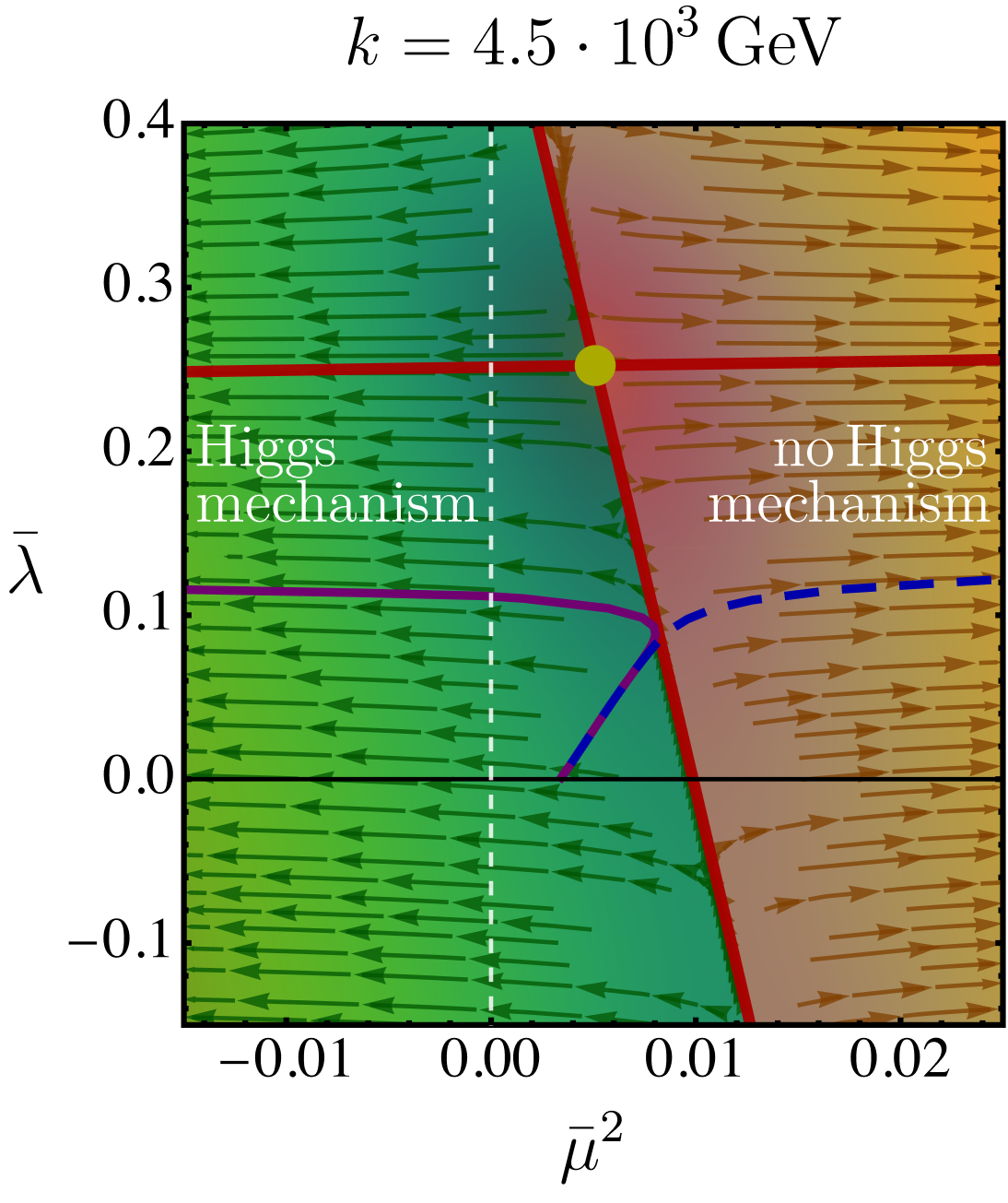}\hspace{0cm}
	\raisebox{0.2\height}{	\includegraphics[width=.56\textwidth]{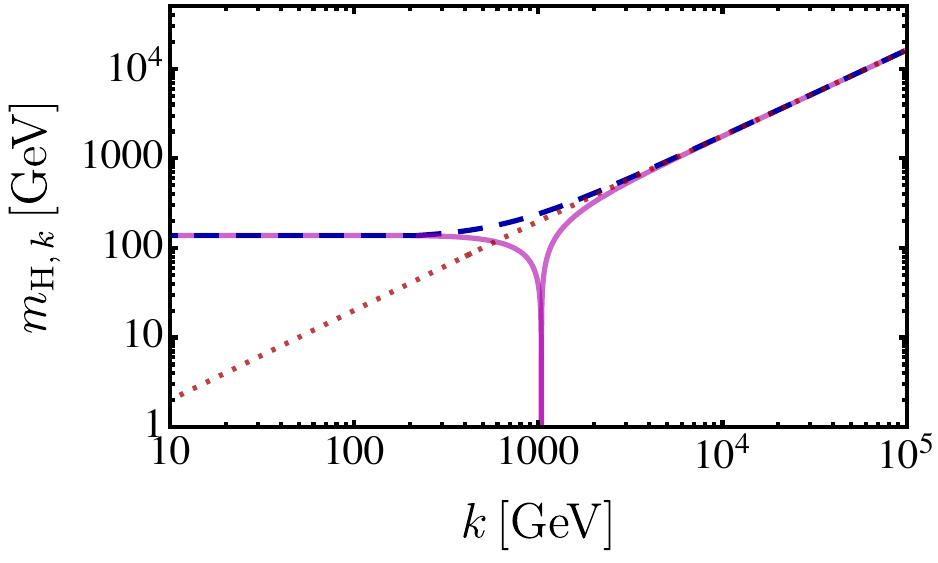}}
	\caption{On the left plot we show the phase diagram for the SM class of theories projected onto the $\bar \mu^2$--$\bar{\lambda}$ plane at a particular scale. We distinguish the SM trajectory (solid purple line) from a neighbouring one with no Higgs mechanism (dashed blue line). The red lines indicate the partial fixed points in \eqref{eq:partialFPmu}, and the vertical dashed white line marks the zero crossing of the curvature mass, signalling the critical scale. The two shaded regions on either side of the separatrix highlight the two possible solutions. On the right plot we show the cutoff dependence of the Higgs Euclidean mass, as defined in \eqref{eq:EuclideanmH}, for the two trajectories depicted in the left plot (same colour coding) and for the massless condition (red dotted) given by \eqref{eq:partialFPmu}.}
	\label{fig:SM phase diagram mHiggs no EWSB}
\end{figure*}

The fact that the SM trajectory lies so \textit{close} to the phase boundary renders it a near (quantum) critical theory. By “close” we refer to the sensitivity of the IR behavior to small variations in the boundary conditions, which could naturally arise from changes in parameters of a UV embedding. Such variations can lead to either the Higgs mechanism occurring or not. Notably, if new physics becomes dynamical around the 10\,TeV scale, variations at the level of 1\% in the Higgs curvature mass, weak gauge coupling, or top Yukawa are sufficient to drive $v \simeq 0$.

A more detailed picture of this near-critical behavior is presented in \Cref{fig:SM phase diagram mHiggs no EWSB}. The left panel shows the SM trajectory in the plane of scalar couplings (purple solid line), which flows to the broken phase in the IR, and a neighboring trajectory (blue dashed line) that flows to the symmetric phase. Both are initialized at $\LUV = 10^{10}$\,GeV with nearly identical UV parameters, lying on opposite sides of the critical surface. The right panel shows that, despite similar IR values for the Euclidean Higgs mass, the two trajectories lead to qualitatively different theories: one with EWSB and the other remaining symmetric, i.e., without a Higgs mechanism.

This illustrates that drastically different IR dynamics—either with or without SSB in the Higgs sector—can emerge from small, natural variations of UV parameters. Theories in the symmetric phase, lacking a Higgs condensate, would exhibit massless fermions and electroweak gauge bosons until chiral symmetry breaking is dynamically triggered by the strong coupling. Consequently, the two non-Abelian gauge groups—SU(3)$_c$ and chiral SU(2)$_L$—would confine sequentially, with color confinement occurring first due to the faster running of the strong coupling. In the absence of explicit fermion masses, the strong sector would break the global chiral symmetry SU$(N_f)_R \times$ SU$(N_f)_L$, giving rise to exactly massless Nambu–Goldstone bosons (pions). This condensate would then indirectly generate fermion and EW boson masses, with a spectrum largely independent of flavor. Altogether, the absence of a Higgs $v$ would yield a Universe radically different from the one we observe.

Finally, we also show in \Cref{fig:SM phase diagram mHiggs no EWSB} the scale-invariant trajectory satisfying \eqref{eq:partialFPmu} (red dotted line). The relative distance between the SM and the critical surface is given by $\Delta \bar \mu^2_\textrm{rel}(\LUV) \approx 0.02 \left( \kSSBone/\LUV \right)^2$.

The near criticality of the SM, in the sense of extreme proximity to theories without Higgs mechanism (e.g. scale invariant), is an intriguing aspect which is not commonly investigated nor used as motivation for model building and we will explore it further in the next section. However, some investigations along this line have been conducted in~\cite{Bellazzini:2015cgj} and in~\cite{Agrawal:1997gf} the context of the anthropic principle.

\section{Quantifying naturalness and the hierarchy problem}\label{sec:naturalness}
In \Cref{sec:SMPT,sec:quantumPT}, we observed how all trajectories with a significant separation of scales are rapidly attracted towards the critical surface in the plane of the Higgs curvature mass and quartic coupling. This clustering of theories in the UV on the critical surface is an inherent feature in the presence of canonically relevant operators, in this case the Higgs mass. As a result, in \eqref{eq:QPTscaling} we see the strong sensitivity of $v$ to the UV scale and the configuration of UV parameters.

In the present Wilsonian RG, where we assume the SM to be a good low-energy theory, the specification of boundary conditions at $\LUV$ is equivalent to considering different fundamental embeddings in which the quantum fluctuations of heavy new physics have been integrated out. For the reader not familiar with this approach, also common to the SMEFT, in \Cref{app:scalarsinglet} we explain in detail how different UV embeddings (new physics) appear as different boundary conditions of the flow at $\LUV$, using a concrete example with a heavy scalar. In summary, $\LUV$ can be roughly understood as the mass scale of the heavy scalar $m_S$, and varying the portal coupling $\bar\lambda_{HS}$ translates into different values of $\Delta \bar{\mu}^2_{\textrm{SM}}(\LUV)$. Consequently, from \eqref{eq:QPTscaling}, it is clear that the higher the scale of new physics becomes, the greater the precision and fine-tuning required to recover the observed IR from the more fundamental configuration. This example allows to interpret the \textit{hierarchy problem} as the difficulty for a UV theory to consistently yield the SM as an EFT.

In this Section, we provide a comprehensive and systematic analysis of the fine-tuning within this class of theories. Various approaches have been used in the literature to quantify fine-tuning, and different measures have been proposed~\cite{Barbieri:1987fn,Giudice:2008bi,Jaeckel:2015txa,Krajewski:2014vea,Ellis:1986yg,Ghilencea:2012gz}. Nevertheless, it is important to highlight that such quantity is not universally defined, and therefore different expressions can be found with same qualitative features and mild quantiative variations. In this work, we will employ the Barbieri-Giudice (BG) measure~\cite{Barbieri:1987fn,Giudice:2008bi} because of its simplicity and frequent use in the high-energy physics community. It is defined as
\begin{align}\label{eq:BGnaturalnessmeasure}
	\Delta_\textrm{BG}= \textrm{max}_{i\,j}\left| \frac{\partial\,\textrm{log}\,{\cal O}_i}{\partial \,\textrm{log}\, g_{j,\,\LUV}}\right|\,,
\end{align}
where ${\cal O}_i$ represents the set of independent observables and $g_{j,\LUV}$ corresponds to the values of the couplings or parameters (Yukawa, gauge, ...) of the EFT at the UV scale.

\begin{figure*}[t!]
	\centering
	\includegraphics[width=.32\textwidth]{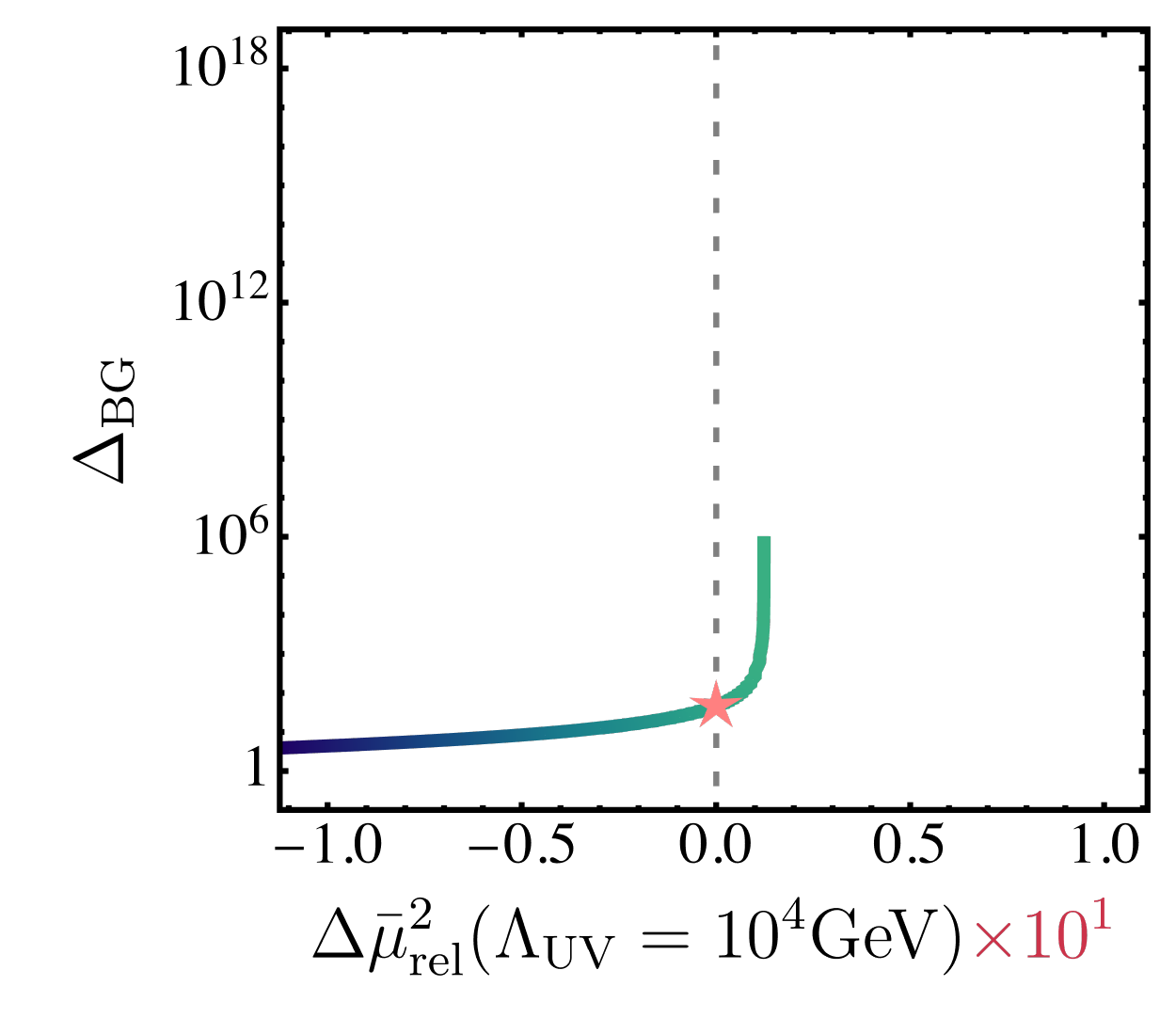}
	\includegraphics[width=.32\textwidth]{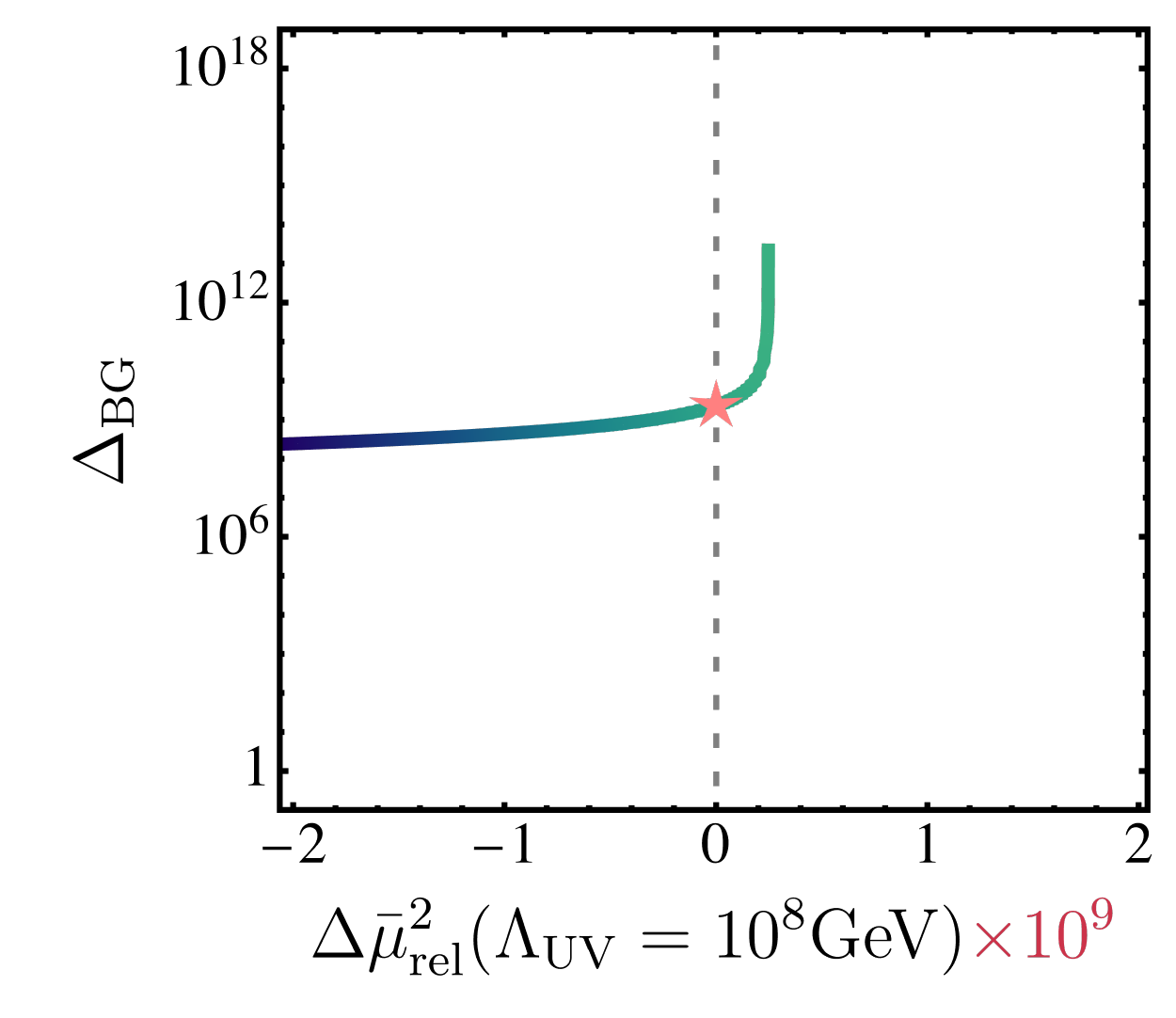}
	\includegraphics[width=.32\textwidth]{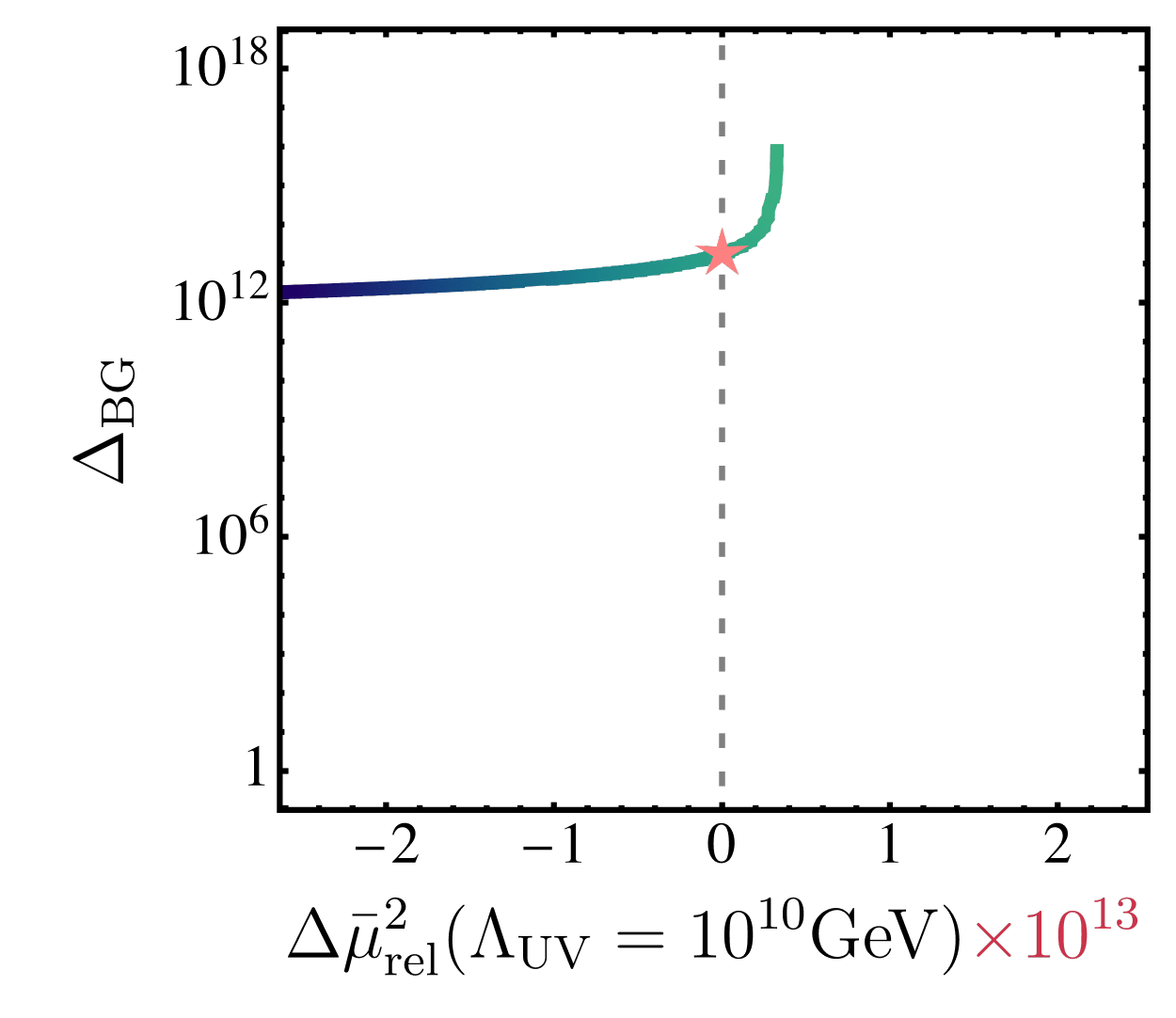}
	\caption{The BG tuning measure as a function of the relative variation of the UV boundary conditions with respect to the physical trajectory reproducing the SM (vertical dashed grey line), $\Delta\bar \mu^2_\textrm{rel}(\LUV)$. From left to right, three choices of the UV scale, $\LUV=\{10^4,10^8,10^{10}\}$\,GeV, are shown in correspondence with the trajectories of \Cref{fig:finetuning SM}. }
	\label{fig:FTSM_diffLambda}
\end{figure*}

For simplicity, let us first consider $v$ as the most susceptible IR observable and $\bar \mu^2_{\LUV}$ as the most relevant parameter at the UV scale. In \Cref{fig:FTSM_diffLambda} we show the BG measure as a function of variations relative to the SM value, \eqref{eq:relativemu0}, for different choices of $\LUV=\{10^4,10^8,10^{10}\}$\,GeV. The BG parameter for theories far from the critical surface is of order of the ratio between scales,
\begin{align}\label{eq:ratioofscales}
(\Delta \Lambda)\equiv (\LUV/1{\rm TeV})\,.
\end{align}
For example, for a UV scale $\LUV=10^{10}$\,GeV (right-most panel in \Cref{fig:FTSM_diffLambda}), $(\Delta \Lambda)^2\approx 10^{14}$ as the curvature mass is relatively varied $\Delta \bar \mu^2_\textrm{rel} (\LUV)\approx(\Delta \Lambda)^{-2}$. If $\LUV$ is considered in the 10 TeV regime, see left-most panel, the fine-tuning required is compatible with that of a theory described purely by marginal operators, e.g. a Yang-Mills theory where one tunes a glueball mass with a separation of scales of $(\Delta \Lambda)\approx 10^{18}$ with $\Delta_\textrm{BG}\approx{\cal O}(10)$~\cite{PastorGutierrez:2025zvx}. Then a separation of scales of up to $(\Delta \Lambda)\approx 10^2$ with $\Delta\bar{\mu}^2_{\rm rel}\lesssim0$ still renders a natural solution with an elementary scalar field. For this reason, the naturalness principle has provided motivation for new physics in the $1\sim10$\,TeV regime, potentially testable by collider-based efforts.  

\begin{figure*}
	\centering
	\includegraphics[width=.375\textwidth]{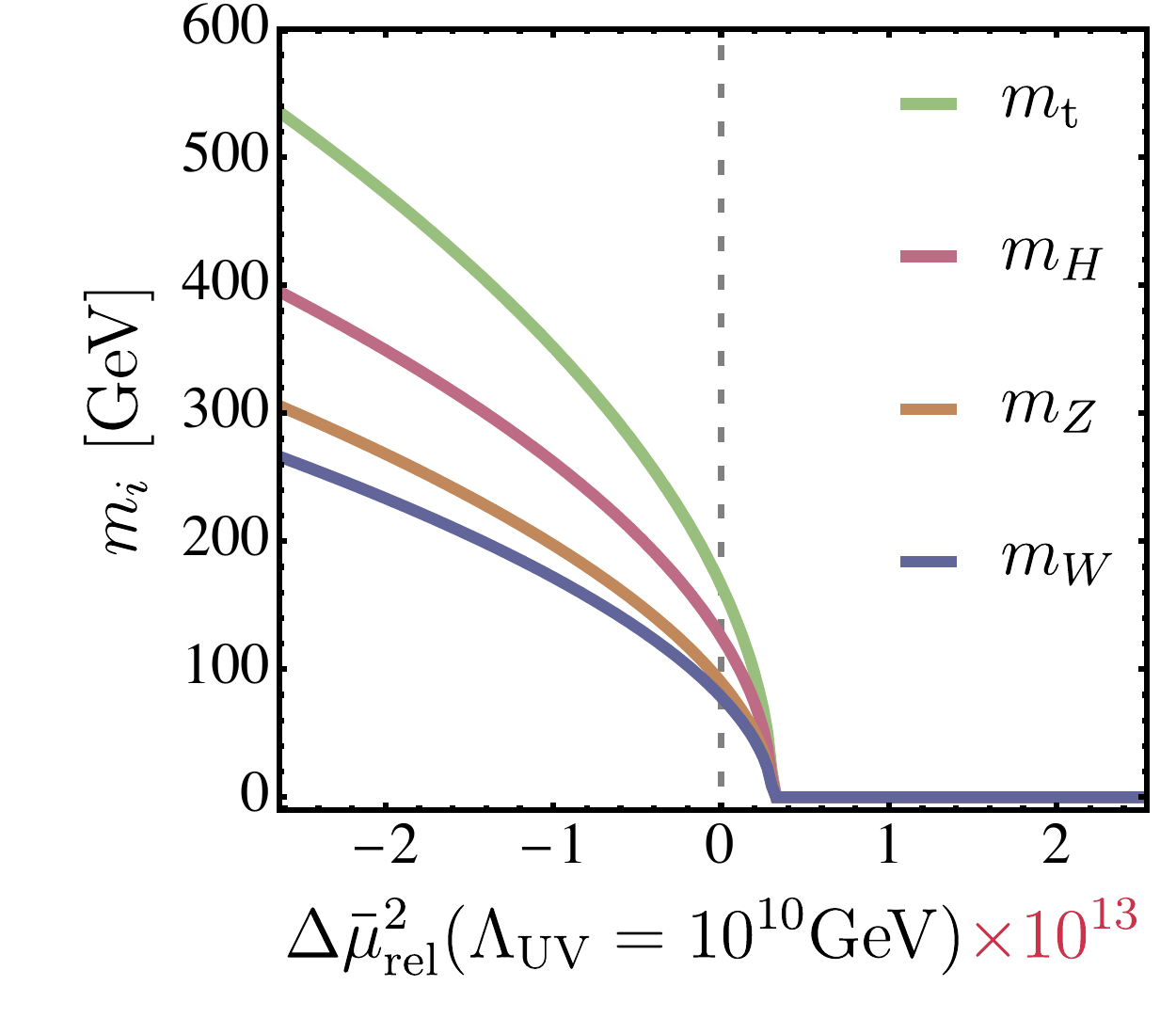}\hspace{.5cm}
	\includegraphics[width=.57\textwidth]{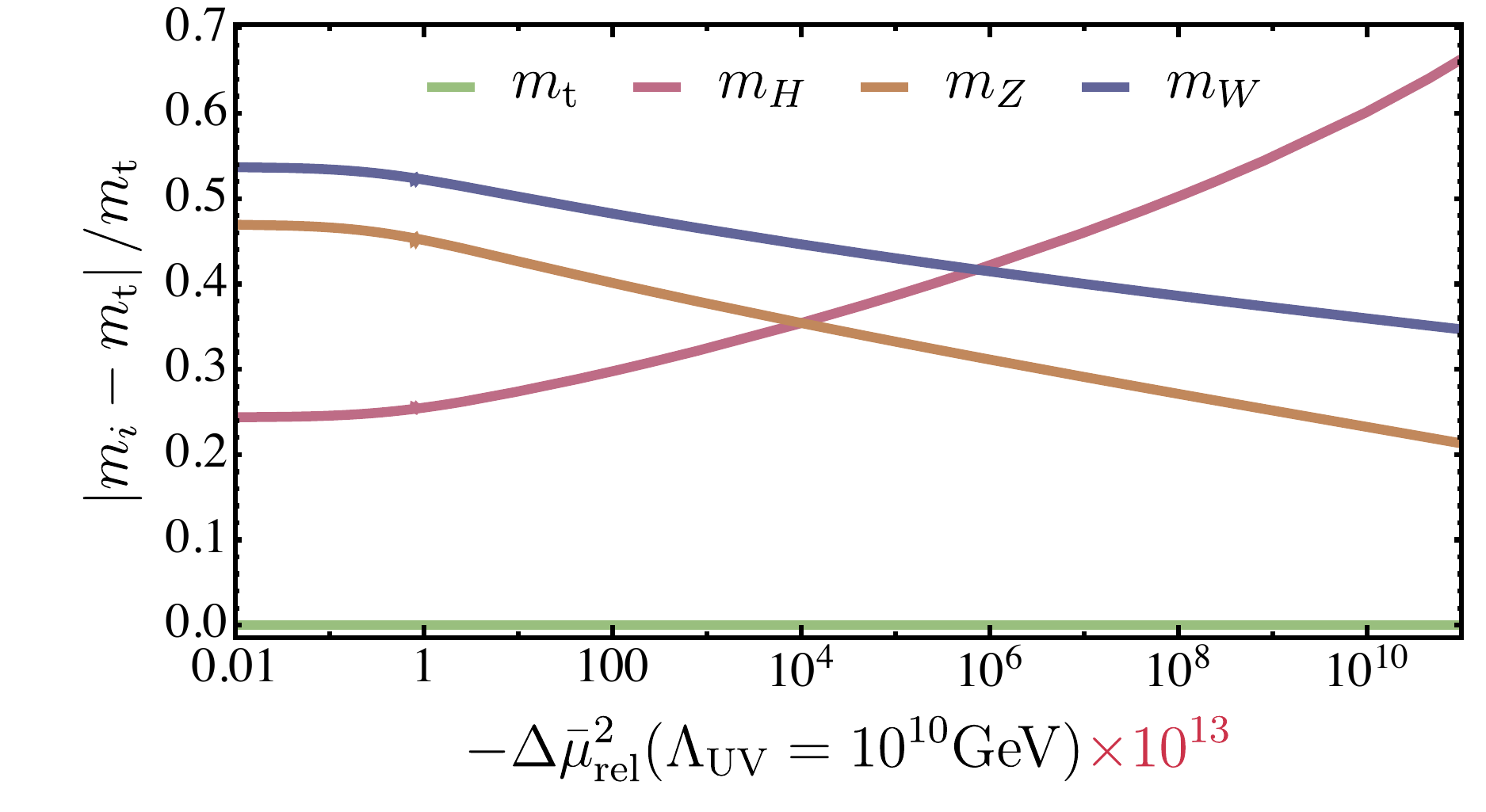}
	\caption{On the left plot we show the Euclidean masses of the top quark, Higgs and electroweak bosons as a function of $\Delta \bar \mu_\textrm{rel}^2 (\LUV)$. On the right plot we show the relative difference between the particle masses with respect to the top quark mass for different $\Delta \bar \mu_\textrm{rel}^2 (\LUV)$. Here, we consider relative variations up to $\Delta \bar \mu_\textrm{rel}^2 (\LUV) \leq 10^{-3}\%$  which suffice to show how the spectrum of the SM particles qualitatively changes.}
	\label{fig:massesFT}
\end{figure*}

\Cref{fig:FTSM_diffLambda} highlights two dominant sources of fine-tuning: the magnitude of the new physics scale ($\LUV$) and the distance of the theory from the critical surface. While the former is well known, the latter is often overlooked, as it requires identifying the presence and structure of the critical surface, making it less apparent. However, this second source induces a sensitivity that can exceed the usual quadratic dependence on $\LUV$ in theories near a quantum phase transition, rendering it highly relevant for a proper quantification of the fine-tuning, in particular in the context of the little hierarchy problem. This dependence is governed by the universal scaling behavior in the vicinity of the massless critical theory. In order to accurately describe the different scalings (particularly with respect to marginal parameters), both power-law and logarithmic corrections have to be taken into account. This is naturally achieved within the present RG approach.

It is important to emphasize, as discussed at the end of \Cref{sec:quantumPT}, that both sources of fine-tuning are scheme-independent, as they depend only on the distance between theories in the UV (which corresponds to reduced quantities in the language of critical phenomena) and not on the precise location of the critical surface or on the regularization procedure~\cite{Aoki:2012xs,Wetterich:2019qzx,Yamada:2020bqe}. In other words, the same physical information can be extracted even within mass-independent schemes. This statement, however, does not preclude the fact that different schemes will, in practice, determine quantities with finite precision.
	
In other mass-dependent regularization schemes, such as those involving a hard UV cutoff, the same qualitative profile as presented here will emerge due to their agnostic character with respect to the UV completion. In contrast, mass-independent schemes such as dimensional regularization require the specification of the UV theory to manifest the threshold and thus capture the fine-tuning specific to a given model. As a result, mass-dependent schemes are useful to explore the landscape of possibilities.

Acknowledging the additional tuning that arises in the near-critical regime (close to the quantum phase transition) we can minimize the overall fine-tuning while keeping the new-physics scale fixed. For this reason, this information becomes valuable for the construction of models that address the problem of little hierarchy with $\Lambda_{\rm UV}\gtrsim 10$\,TeV.

\begin{figure*}[t!]
 	\centering
 	\includegraphics[width=.75\textwidth]{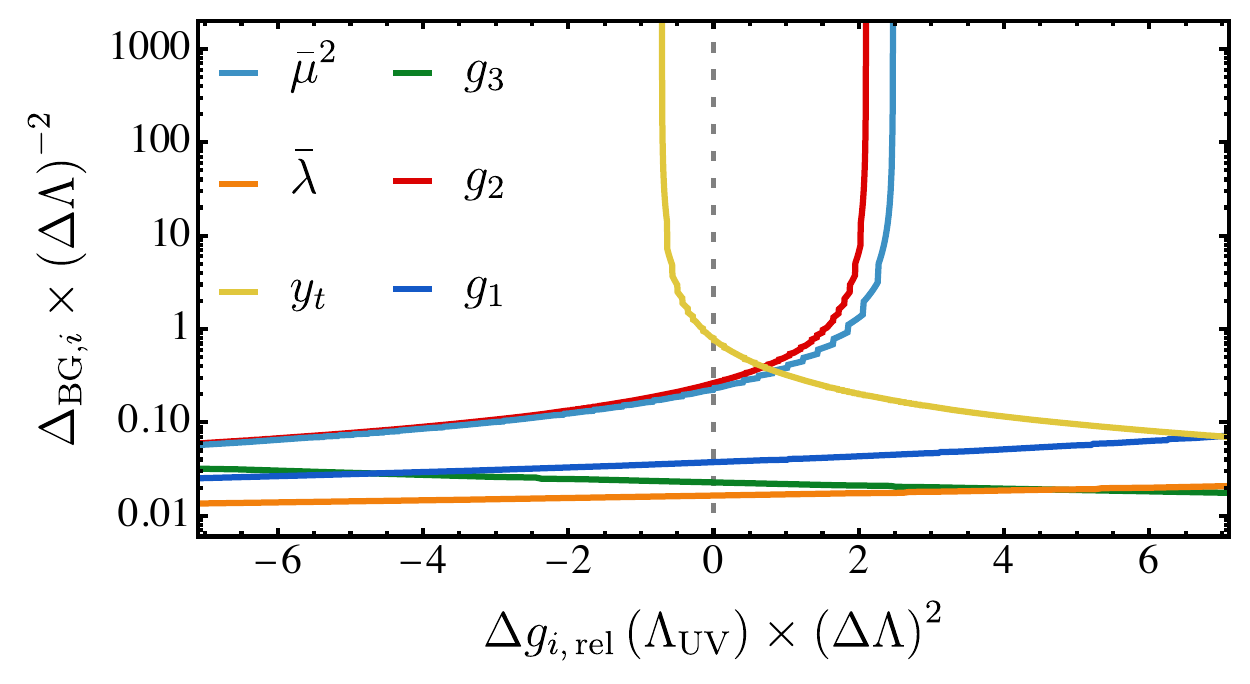}
 	\caption{ Dimensionless BG measure as a function of the relative variation of the different SM parameters at an arbitrary UV scale $\LUV= \Delta \Lambda \cdot 1$\,TeV. }
 	\label{fig:FTSM_diffgi}
\end{figure*}

The BG measure, as defined in \eqref{eq:BGnaturalnessmeasure}, identifies the most sensitive direction in theory space as the one determining the degree of fine-tuning. In the results shown in \Cref{fig:FTSM_diffLambda}, we considered the variations only in $\bar\mu^2$ as the single UV parameter and used $v$ as the only observable IR observable. We now extend the analysis to explore other directions in UV theory space, as well as additional IR observables. In the left panel of \Cref{fig:massesFT}, we display the mass spectrum of the EW gauge bosons, Higgs, and top quark as a function of relative variations in the curvature mass at $\LUV = 10^{10}\,$GeV. For theories in close proximity to the SM, the sensitivity of these IR observables is primarily driven by variations in $v$, thereby justifying the earlier approximation of neglecting the variation of other couplings.

However, as shown in the right panel of \Cref{fig:massesFT}, for larger UV deformations (e.g., curvature mass variations of order $0.0001\%$), shifts in the marginal couplings do induce observable changes in the mass hierarchies — in some cases leading to scenarios where the Higgs becomes lighter than the EW gauge bosons. However, in the following, we restrict our analysis to theories in the near vicinity of the SM and continue to use $v$ as the most sensitive IR observable for computing the BG measure.

We now proceed to investigate the sensitivity of IR observables to variations along other directions in theory space beyond the curvature mass, such as the gauge and Yukawa couplings. From the perspective of the EFT, these variations can be linked to integrating out different types of new physics in the respective sectors. For example, new physics coupled predominantly to the top quark will manifest itself in the quadratic sensitivity via the second term in square brackets in \eqref{eq:power-lawflowmu}. In \Cref{fig:FTSM_diffgi}, we present the dimensionless version of the BG measure, where the amount of fine-tuning has been normalized by $(\Delta \Lambda)^2$ to remove the leading canonical scaling. The Figure shows the dependence of this normalized measure on relative variations in different directions in theory space. The trajectories shown were initiated at $\LUV = 10^8\,$GeV, and only slight differences are observed for other choices of the UV cutoff scale.

The IR observables turn out to be most sensitive to variations in the top Yukawa coupling, followed by $g_2$, and then by $\bar{\mu}^2$. This observation is well understood in terms of the mixing between the canonically relevant curvature mass and the marginal couplings. In particular, the Yukawa and gauge couplings provide dominant contributions to the flow of $\bar{\mu}^2$, as evident from \eqref{eq:flowmu}. This relevant mixing is also reflected in the off-diagonal components of the stability matrix.

We emphasize that the important idea to have in mind when interpreting the results obtained in this Section is that the effects of heavy degrees will be such that a change on their associated parameters relative to those rendering the SM in the IR can be cast as shifts in the effective SM boundary conditions, see \Cref{app:scalarsinglet}. In this way, the above results give information about how sensitive the effective SM is to such shifts in the UV embedding parameters. Furthermore, dialing fundamental parameters of different UV embeddings will generally affect the effective SM boundary conditions in different amounts and in different combinations. Since we remain agnostic about the particular nature of the heavy degrees of freedom, the above results explore shifts in all relevant directions of the SM.

Finally, we would like to discuss the interpretation of  \Cref{fig:FTSM_diffgi}: reducing the top Yukawa renders theories closer to the critical surface and hence enhances the tuning. This is well understood as fermionic contributions are known to enhance SSB, therefore reducing those renders the realisation of a Higgs mechanism more challenging. Nonetheless, note that this is obtained considering that all other parameters in the theory remain fixed (and allowing no BSM physics below $\LUV$). This would however lead to a theory differing from the SM in the IR, notably with a smaller Higgs boson mass. In the view of model building and phenomenologically viable BSM scenarios, this would need to be counter-balanced by further changes to the theory (see also \Cref{sec:newPhysicsdeformations}), in particular to keep the Higgs mass at its observed finite value away from criticality -- so one cannot generically conclude that BSM models with a smaller top Yukawa are more tuned. Still, \Cref{fig:FTSM_diffgi} highlights the generically enhanced tuning in the top Yukawa that we find due to the {\it vicinity to criticality}, increasing the naive tuning considerably.

\section{Phase diagram deformations and new physics}\label{sec:newPhysicsdeformations}
\begin{figure*}[t]
	\centering
	\includegraphics[width=.8\textwidth]{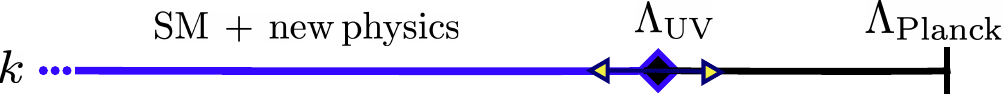}
	\caption{Sketch of the approach to new physics followed in this Section. We consider that some form of new physics coexists with the SM, acting as a viable EFT below an unknown $\LUV$ scale.
	}
	\label{fig:SM+new}
\end{figure*}

We continue by discussing how the developed framework could be used to formulate new BSM directions that alleviate the IR-UV sensitivity present in the SM as an EFT. So far, we have followed a hard cutoff approach that mimics the presence of new physics beyond a certain UV scale. Consequently, we have remained completely general and agnostic with respect to the nature of the new physics entering above the UV cutoff scale, as the Wilsonian approach does not require specifying a UV embedding to manifest the quadratic sensitivity of the low-energy physics to UV scales. In the following, we assume that below the considered $\LUV$ scale, the SM is extended by some new physics coupled only to the Higgs sector. \Cref{fig:SM+new} summarizes the strategy followed in this section. The main goal of this analysis is to investigate the possibility of non-trivial deformations of the SM that screen the sensitivity to other UV physics. In other words, we explore which modifications --such as changes in matter content or couplings-- should be implemented in the EFT to balance the increase of the UV scale, as required by pressing experimental constraints. 

Under the above assumptions, the new-physics effects on the dynamics of the theory can be captured by modifying the RG flow of the scalar couplings to
\begin{align}
	&\partial_{t} \bar \mu^2 = \left(\partial_{t} \bar \mu^2\right)_\textrm{SM} + c_1 \cdot \bar \mu^2 + c_2 &&\textrm{and}&&&\partial_{t} \bar \lambda = \left(\partial_{t} \bar \lambda\right)_\textrm{SM} + c_3 \cdot \bar \lambda + c_4\,. \label{eq:flowdeformations}
\end{align}
Here, $\left(\partial_{t} \bar \mu^2\right)_\textrm{SM}$ and $\left(\partial_{t} \bar  \lambda\right)_\textrm{SM}$ stand for the RG flows of the SM class of theories, and the coefficients $c_i$ parameterize the contributions of new physics. The dependence of the set of coefficients $c_i$ on new-physics interactions, such as those given by a new Yukawa coupling $y_N$, a new gauge coupling $g_N$ or a new scalar portal $\bar \lambda_{HS}$, all involving only interactions between new particles and the Higgs boson, has the form
\begin{align}
	& c_1 = f_1\left(-g_N^2,\, y_N^2,\,\ldots\right), && &&& c_2 = f_2\left(-\bar\lambda_{HS},\,-g_N^2,\, y_N^2,\,\ldots\right),\notag\\[1ex]
	& c_3 = f_3\left(-g_N^2,\, y_N^2,\,\ldots\right), &&\textrm{and}&&& c_4 = f_4\left(\bar\lambda_{HS}^2,\, g_N^4,\,-y_N^4,\,\ldots\right)\,,\label{eq:deformationsandBSM}
\end{align}
where $f_i$  are the multi-loop functions encoding the symmetry coefficients and the strength of the coupling to the new sector. Each $c_i$ affects the flows of the scalar sector in a different way. More specifically, $c_1$ alters the contributions to the anomalous dimensions of the Higgs field, $c_2$ the power-law part of the flow, and $c_3$ and $c_4$ the flow of $\bar\lambda$. Evidently, when considering some specific new physics, the relation between several deformations $\{c_i\}$ will be given uniquely. However, in the present stage of the analysis, we keep them free. Additionally, we further simplify the setup by neglecting the flow of the underlying couplings in $c_i$. It is evident that the present approach aims for a qualitative understanding of the possible deformations of theory space, and a more quantitative implementation is needed for a quantitative assessment.

In the following, we identify the physical theory as the one that reproduces the SM values at $\kSSBone$, namely $\bar \lambda_{k\to \kSSBone} \approx 0.13$. Regarding the rest of the SM gauge couplings and Yukawas, as they do not sense significant modifications to their previously fixed IR boundary conditions, we neglect their variations in the IR.  Lastly, we do not enter the broken phase and instead read the value of $v$ from $\kSSBone$. The existence of such mapping is clear from the comparison of the last two columns in \Cref{fig:finetuning SM}.

\begin{figure*}[t!]
	\centering
	\includegraphics[width=.55\textwidth]{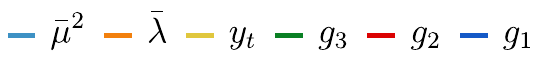}\\[2ex]
	\includegraphics[width=.45\textwidth]{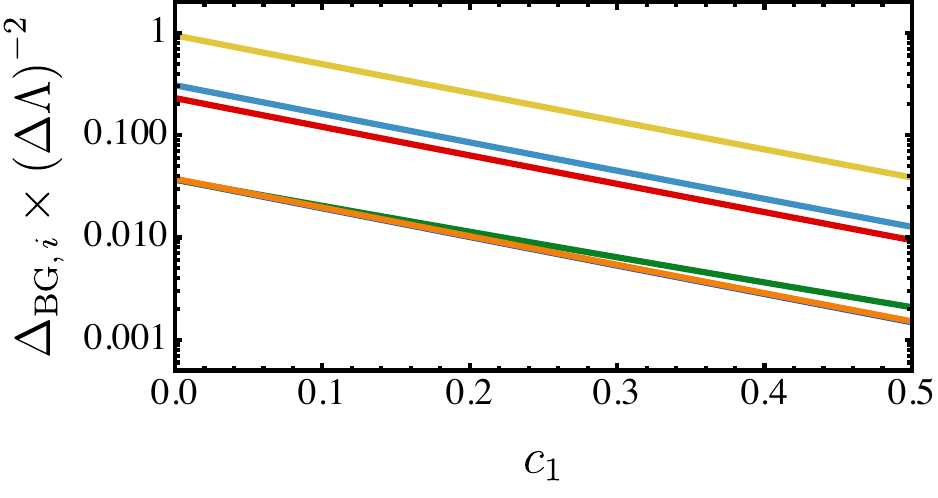}\hspace{1cm}
	\includegraphics[width=.45\textwidth]{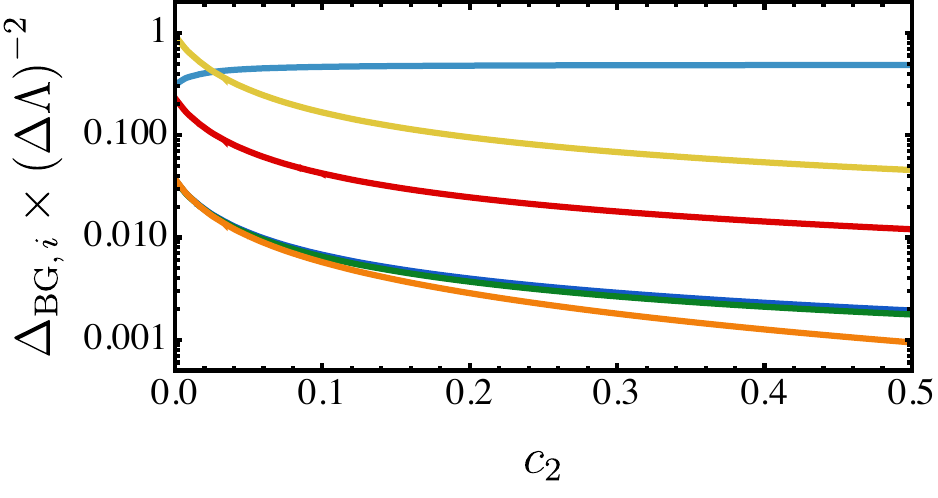}\\[2ex]
	\includegraphics[width=.45\textwidth]{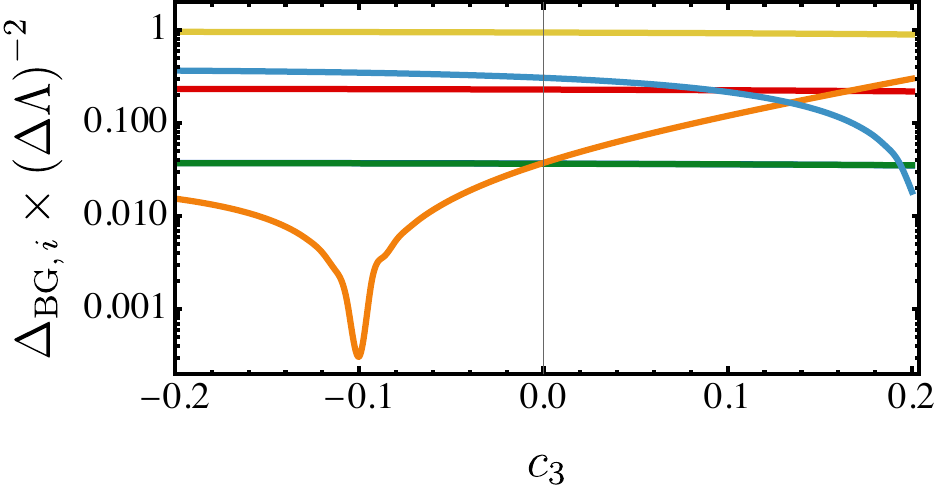}\hspace{1cm}
	\includegraphics[width=.45\textwidth]{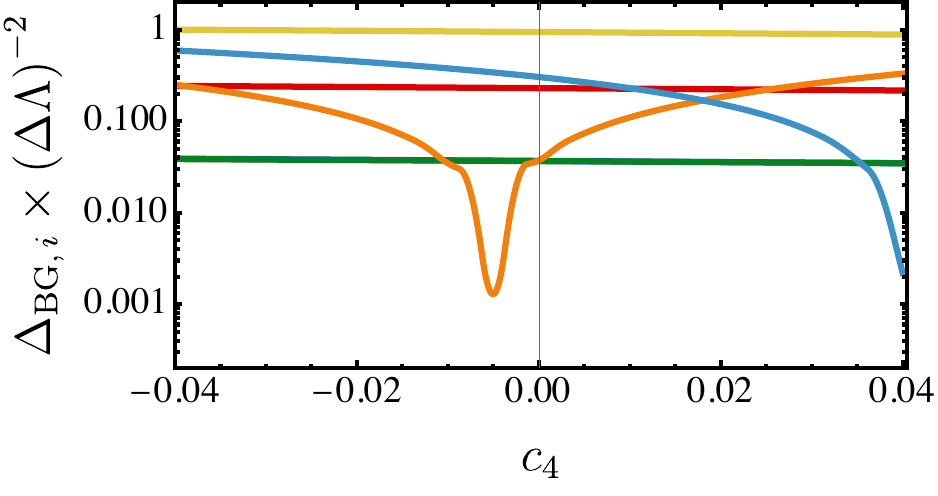}
	\caption{In the different pannels we show the dimensionless version of the fine-tuning parameter, $\Delta_{\textrm{BG},i} \times (\Delta \Lambda)^{-2}$, as a function of the deformations introduced in \eqref{eq:flowdeformations} which parametrize the inclusion of new physics, see \eqref{eq:deformationsandBSM}. The coloured lines show the sensitivity along variations in different directions of theory space, namely in the various couplings.
	}
	\label{fig:DBG_ci}
\end{figure*}

In \Cref{fig:DBG_ci} we display the BG measures for trajectories compatible with the SM in the IR and for different directions of the theory space. The effects of new physics are parametrized by the different $c_i$'s which we dial independently in this simplistic exercise. The final $\Delta_\textrm{BG}$ is determined as the maximum of all variations in theory space, which corresponds to the highest of the curves. Although the results were obtained for $\LUV = 10^6$\,GeV, we show the rescaled version of the BG measure where we divide by $\left(\Delta \Lambda\right)^2 $ defined in \eqref{eq:ratioofscales}. Varying the UV cut-off has a minor impact on the $c_i$-dependence, the overall picture remains the same.

In the upper-left panel, we observe that the highest sensitivity arises when varying the top Yukawa, as previewed in \Cref{fig:FTSM_diffgi}. However, there is a constant decrease in the fine-tuning in all directions of theory space as $c_1$ increases. This parameter enhances the Higgs anomalous dimension and effectively decreases the dimensionality of the Higgs mass operator. In other words, such deformation renders the curvature mass operator more canonically irrelevant at $\LUV$ and, as expected, reduces the level of fine-tuning. 

In the top-right panel we vary $c_2$ which corresponds to a correction to the power-law contribution to the flow of the curvature mass. Such a deformation could be generated from bosonic tadpole diagrams, which do not affect the anomalous dimension at leading order because of their momentum independence. For large values of $c_2$, we observe how the fine-tuning in the direction of the marginal couplings decreases, while it mildly increases and eventually freezes in the direction of the curvature mass. In combination, we see only a very mild improvement in fine-tuning. Negative variations of $c_2$ are also allowed, and these displace the partial fixed point in the $\bar \mu^2-\bar\lambda$-plane towards the regime with a nontrivially stable shape. However, the resulting trajectories develop a nontrivially stable shape at high energies. This is the same scenario as the one presented in~\cite{Goertz:2023pvn} and discussed in other contexts~\cite{Hamada:2012bp,Jones:2013aua,Holthausen:2011aa}. Given that this requires a more delicate treatment due to the presence of a flowing minimum and potential decoupling of modes, we do not discuss it here.

In the bottom panels, we vary $c_3$ (left panel) and $c_4$ (right). Positive deformations of these parameters lower the position of the horizontal partial fixed point. In fact, this triggers the stabilization of the quartic coupling towards the UV and ultimately lowers the associated Landau pole. These deformations are interesting because they improve or even cure the metastability of the Higgs sector. Also, they reduce the fine-tuning in the direction of the Higgs parameters. This minimization occurs at two instances. First, at the maximum values of the deformations shown, $c_3\simeq0.2$ and $c_4\simeq 0.04$, which correspond to the scenarios in which the trajectories displaying the SM IR values have a stable shape and $\bar\mu^2_{\LUV}=0$ (again entering the same class as that discussed in~\cite{Goertz:2023pvn}). The second instance occurs for $c_3\simeq-0.1$ and $c_4\simeq -0.005$, and corresponds to the case where the quartic coupling vanishes at the UV scale. However, it is important to note that the sensitivity with respect to other (marginal) coupling directions remains rather constant. As the dominant sensitivity is still sourced in the Yukawa direction, which is rather insensitive to these deformations, the total BG measure remains unchanged and no fine-tuning is relaxed. However, we emphasize that in more realistic model embeddings, where specific combinations of $c_i$ are required, the dependence on deformations could differ, potentially revealing directions that reduce the overall fine-tuning. This is in particular also true for scenarios that couple the new physics differently to the different SM sectors, which calls for further investigation. Especially in the context of the little hierarchy problem, where a mild improvement is sufficient, more natural models could be found.

In summary, we observe that it is not easy to find more natural configurations in theory space via the simple corrections considered here and in the presence of a common UV scale, other than by reducing the canonical dimension of the Higgs mass operator~\cite{Aoki:2012xs}. The dominant sensitivity in the marginal coupling directions arises as a direct consequence of the mixing between the relevant direction and the marginal couplings. The most promising pathway appears to be the simplest one: introducing new physics that effectively lowers the dimension of the Higgs mass operator. In fact, this has been explored in the context of \textit{self-organisation}~\cite{Bornholdt:1992up,Gies:2025pqv,Wetterich:2016uxm}. Also, this direction includes models where the Higgs field operators can be formulated in a more fundamental way from higher-order fermionic interactions; see e.g.~\cite{Kaplan:1983fs,Kaplan:1983sm,Dugan:1984hq,Hill:2002ap,Agashe:2004rs,Cacciapaglia:2020kgq,Cacciapaglia2022a} 
for Composite Higgs models and~\cite{Ahmed:2023qsm} for similar recent proposals which complete Little Higgs models in a UV-safe way, as well as~\cite{Goertz:2023nii,Goertz:2024dnz} for treatments within a first-principles approach. Other interesting implementations in this direction could be achieved by coupling an elementary Higgs to a strong sector or by considering a large multiplicity of new dark fields~\cite{Gies:2025pqv}. All these directions benefit from a non-perturbative treatment of the new-physics scenario, which will be explored in more detail in an independent work. 
 
The present investigation also allows us to identify specific configurations that merit further discussion. One corresponds to the solution for which the power law part of the flow exactly cancels at the UV scale, $\left.\partial_{\bar \rho} \,\,\overline{\text{Flow}}\left[V_{\textrm{eff}}\right]\right|_{\bar \rho_0=0}=0$ in \eqref{eq:flowmu}. In \Cref{fig:DBG_ci}, this occurs for $c_2 \simeq -10^{-2}$, $c_3 \simeq 0.2$, and $c_4 \simeq 4\cdot 10^{-2}$. This particular solution is known as the \textit{Veltman condition}~\cite{Veltman:1980mj}, which gained relevance in high-scale SUSY extensions, as it determines the corresponding symmetry-breaking scale. Above such, the power-law corrections are ensured to be zero given the existing symmetry. Other interpretations for the energy scale at which the Veltman condition is satisfied have been proposed; see~\cite{Hamada:2012bp,Jones:2013aua,Holthausen:2011aa} for examples related to the appearance of quantum gravity and~\cite{Abu-Ajamieh:2021vnh,Masina:2013wja} for applications in BSM extensions. In~\cite{Goertz:2023pvn} it has been shown how the change in the sign in the power-law correction, in fact, reflects a change in the curvature of the effective potential and leads to the emergence of non-trivial UV phases of the Higgs potential. In relation to the hierarchy problem, as discussed before the Veltman Condition does not reduce the overall tuning but only in one direction of theory space. Moreover, we note that the present EFT approach ignores the presence of multiple particle thresholds which is a relevant quality for the satisfaction of the condition in realistic proposals~\cite{Lindner:1992ah}. In fact, in \eqref{eq:power-lawflowmu} we assumed a universal BSM scale coupled to the scalar, fermion, and gauge sectors.
 
Another particular solution is the one that places the theory on the partial fixed point, i.e. the trajectory for which the \textit{total} flow of $\bar\mu^2$ vanishes at all scales. As discussed in \Cref{sec:quantumPT}, this is known as the massless or critical condition~\cite{Aoki:2012xs,Wetterich:2019qzx,Wetterich:1983bi}, where the Euclidean Higgs mass satisfies $m^2_{H,k\to0} \simeq 0$. This solution effectively realizes scale symmetry, as no mass scale is generated at $k \to 0$, neither via SSB nor through an explicit Higgs mass term. In the present simplified analysis, this solution is satisfied only at the $\LUV$ scale for parameter values that approximately also satisfy the Veltman condition.  At lower scales, the theory flows away from the partial fixed point along a marginally relevant direction in theory space. To achieve a fully scale-invariant solution, the massless condition in \eqref{eq:partialFPmu} needs to be satisfied for all scales and only broken spontaneously.

\section{Conclusions}\label{sec:conclusions}

Naturalness of the electroweak scale provides one of the main guiding principles in the search for a more fundamental realization of nature beyond the Standard Model (SM), and it is inherently rooted in the quantum corrections to the Higgs mass. In order to mitigate the large quantum corrections which come along with a severe fine-tuning, new physics is generally expected to appear at scales of order~TeV. 
The absence of new physics at the LHC suggests that the scale of new physics should be higher, leading to the so-called little hierarchy problem. As a result, proposed solutions to the big hierarchy problem face additional constraints, since the generic new physics scale needs to be raised while still remaining natural (or very different mechanisms need to be considered). Most approaches along these lines rely on mechanisms that postpone the full impact of the hierarchy problem. There might or might not be a fundamental link between such mechanisms and the physics invoked to address the full hierarchy problem, though such a connection might be a hint towards a consistent solution.

To add to the understanding of these pressing questions, in this work we revisited the notion of fine-tuning present in the hierarchy problem from the viewpoint of critical phenomena. For this, we have studied the phase structure of the SM class of theories using the functional Renormalization Group (fRG), a Wilsonian and mass-dependent renormalization group scheme that offers a distinct perspective to the common perturbative approaches. The developed framework allowed deepening the understanding of the hierarchy problem by emphasizing that SM-like theories with a new scale are not only quadratically fine-tuned in the conventional sense, but also near quantum critical. We stressed that, extrapolating the low energy description up to some large new physics scale not only imposes conditions on quadratic corrections, but logarithmic contributions must also be considered in the tuning, which is relevant for UV completions of the SM. In particular, we showed that the near criticality implies an enhanced tuning in the neighborhood of the SM theory space that should be acknowledged in naturalness studies, especially regarding the little hierarchy problem. 

In this paper we first reviewed the RG flow of the Higgs mass and its implications on the theory's phase diagram and special trajectories. The IR regularization intrinsic to the fRG, together with its ability to account for power-law running, allowed us to track the evolution of the effective Higgs potential along the RG-flow, akin to a finite-temperature system. This feature enabled a detailed analysis of the phase diagram, including the identification of the critical surface and associated distinct trajectories (e.g. the scale invariant one), in a intuitive and graphical way.

We then investigated the properties of the quantum phase transition that separates theories with and without a Higgs mechanism. The associated critical behavior is characterized by universal quantities such as critical exponents, which we linked directly to standard measures of fine-tuning. Notably, we emphasized the conceptual connection between the proximity of the SM to criticality and its sensitivity to UV physics — two phenomena that, while distinct, share a common physical origin in the presence of new physics. This connection proves particularly useful in the context of naturalness, as it allows one to formulate the degree of tuning in a physical and scheme-independent manner. We have stressed that, within this finite RG approach — where no divergences are encountered — the IR–UV sensitivity can be expressed in terms of well-defined and universally derivable quantities. This stands in contrast with some interpretations in other finite schemes~\cite{Mooij:2024rys,Mooij:2021lbc,Mooij:2021ojy}.

The low-energy effective theory viewpoint adopted here provides a general framework for studying the sensitivity of IR observables to unknown UV completions in an agnostic way. Within this setting, we identified two distinct sources of fine-tuning. First, the well-known tuning due to the separation between the electroweak scale and that of new physics (mimicked by a UV cutoff), and second, an additional source stemming from the theory's proximity to the critical surface. While the former leads to the standard quadratic sensitivity, the latter generates even stronger dependencies, particularly for theories near the quantum phase transition. In fact, some mild heavy new-physics deformations can eliminate the Higgs mechanism entirely, rendering radically different IR spectra. This latter tuning dimension is, in principle, accessible in any scheme and could open new avenues for model building, also in the context of solving the little hierarchy problem.

To illustrate the practical utility of our framework, we presented an example where we search for low-energy extensions only coupled to the Higgs sector which reduce the sensitivity of IR observables to heavy new physics. Within this simplistic analysis, we also recovered well-known mechanisms involving large anomalous dimensions for the Higgs field, thus demonstrating that such scenarios can be naturally embedded and analyzed within our approach.

\begin{acknowledgments}
We would like to thank Holger~Gies, Satoshi~Iso, Joerg~Jaeckel, Prisco~Lo~Chiatto, Jan~M.~Pawlowski, Masatoshi~Yamada and Christof Wetterich for fruitful discussions. We thank Andrei Angelescu for his collaboration in the early stages of the project. JPG acknowledges funding from the International Max Planck Research School for Precision Tests of Fundamental Symmetries (IMPRS-PTFS). APG is supported by the RIKEN Special Postdoctoral Researcher Program (SPDR).
\end{acknowledgments}

\appendix
\section*{Appendices}
\section{EFT approach to new physics with Wilsonian schemes} \label{app:scalarsinglet}
In this Appendix we detail the EFT approach employed to describing heavy new physics and its manifestation in the tuning. In particular, we discuss the well-known identification of new physics on the UV cut-off. The only peculiarity is that here we enjoy a mass-dependent scheme which progressively includes the decoupling effects over a range of scales.

To clarify the mechanism, we consider an explicit example in which the SM is extended by the introduction of a heavy scalar field coupled to the Higgs via a portal coupling $\bar \lambda_{HS}$. At lowest order, this new physics scenario only enters via the power-law correction in \eqref{eq:power-lawflowmu}, and does not sizably affect the anomalous dimension due to the momentum independence of the tadpole diagram which only enters very subdominantly. Therefore, for simplicity, we consider that such corrections only enter to the Higgs mass via
\begin{align}\label{eq:newphysicspowerlaw}
	\left.\partial_{\bar \rho} \,\,\overline{\text{Flow}}\left[	V_{\textrm{eff}}\right]\right|_{\bar \rho_0=0}\supsetsim -\frac{\bar \lambda_{HS}}{8 \pi^2(1+\bar m_S^2)^2} =-\frac{k^4\, \bar \lambda_{HS}}{8 \pi^2(k^2+ m_S^2)^2} \,.
\end{align}
Note that this also neglects the corrections to the Higgs quartic coupling and any other higher-dimensional operator. 

At energy scales approximately below the mass of the scalar $k\lesssim m_S$, the quantum corrections respective to the new physics begin to decouple, becoming progressively less relevant in the off-shell contributions to the SM and the new physics parameters ($m_S$ and $\lambda_{HS}$). This decoupling is given by the smooth and continuous threshold function $(k^2+ m_S^2)^{-2}$ and as we go to lower energy scales the contribution of the new physics becomes weaker and weaker and the vanilla SM becomes a better and better EFT. 

Let us now consider a hard UV cutoff to reproduce the effect of this setup. At the energy scale $k\lesssim m_S$, the contribution to the flow of the mass is $\lesssim -\bar \lambda_{HS} /(4*8 \pi^2)$ and can be reasonably neglected from lower scales on. Therefore, we may put a UV cutoff $\LUV$ at this scale and describe the theory effectively purely as the vanilla SM. Now, different boundary parameters at the UV will match different theories in the IR. For example, having $\LUV=m_S$ fixed, a deviation from the boundary conditions which render the SM is matched to different portal couplings;
\begin{align}
	\Delta \bar \mu^2_\textrm{rel} (\LUV)=-\frac{\bar \lambda_{HS,\, \LUV}-\bar \lambda_{HS,\,\LUV,\,{\rm SM}}}{\bar\lambda_{HS,\,\LUV,\,{\rm SM}}}\,,
\end{align}
where $\bar\lambda_{HS,\,\LUV,\,{\rm SM}}$ denotes the fundamental theory that renders the SM in the IR. Analogously, we should also include such deviations in the variation of the other parameters (e.g., $\Delta\bar \lambda_{\rm rel}$, $\ldots$) but their sensitivity is considerably subdominant, given the larger number of powers in the threshold function. 

What we have just discussed trivially extends to corrections that also enter dominantly via the anomalous dimension. For example, heavy fermions correct the power law and the logarithmic term. Those contributions can also be absorbed in the boundary conditions at the UV boundary.

Although here we have made a number of simplifications and neglected effects such as those from higher dimensional operators or the fact that the threshold function is not a step function at $\LUV$, we state that this approach can be made quantitative as in SMEFT or HEFT with the fRG approach. With this example, it appears clear the power of effective approaches for their generality.

We remind the reader that the hierarchy problem is formulated in a physical manner: \textit{the sensitivity of IR observables to UV physics of a more fundamental embedding}. While we can always put a UV cutoff and check the sensitivity to modifications at those scales, it only gains physical sense when we give physical meaning such as the one just explained with the heavy scalars. Additionally, here we see once again that only the scheme-independent distance between theories is relevant.

\bibliographystyle{JHEP}
\bibliography{references}

\end{document}